\documentclass[12pt,a4paper]{article}
\usepackage{a4,epsf,here,epsfig}
\def\be{\begin{equation}}
\def\ee{\end{equation}}
\def\bea{\begin{eqnarray}}
\def\eea{\end{eqnarray}}
\def\eq#1{(\ref{#1})}
\def\bs{\bigskip}
\def\ms{\medskip}
\def\hom{\hbar\omega}
\def\fig#1{figure \ref{#1}}
\def\etal{{\it et al.}}

\def\siml{\,\hbox{\kern.1em \lower.6ex \hbox{$\sim$} \kern-1.12em
          \raise.6ex \hbox{$<$} }}
\def\simg{\,\hbox{\kern.1em \lower.6ex \hbox{$\sim$} \kern-1.12em
          \raise.6ex \hbox{$>$} }}
\def\d{{\rm d}}

\newcommand{\Figurebb}[9]{
\begin{figure}[H]
\begin{center}
\leavevmode
\epsfysize=#7cm
\epsfbox[#2 #3 #4 #5]{#6}
\par
\parbox{#8cm}{
\caption[figure]{\renewcommand{\baselinestretch}{0.8} \small
                                           \hspace{-0.3truecm}#9}
\label{#1}}
\end{center}
\end{figure}
}

\def\eq#1{(\ref{#1})}

\begin{document}

\centerline{\bf \Large Semiclassical trace formulae for pitchfork bifurcation sequences}

\ms
\bs

\centerline{\bf J. Kaidel$^1$ and M. Brack}

\ms

{\small
\centerline{Institute for Theoretical Physics, University of
Regensburg, D-93040 Regensburg, Germany}
\centerline{$^1$e-mail: joerg.kaidel@physik.uni-regensburg.de}
}
 
\bs

\noindent 
{\bf \large Abstract} 

\ms
\noindent
{\small
In non-integrable Hamiltonian systems with mixed phase space and
discrete symmetries, sequences of pitchfork bifurcations of periodic 
orbits pave the way from integrability to chaos. In extending the
semiclassical trace formula for the spectral density, we develop a uniform 
approximation for the combined contribution of pitchfork bifurcation pairs. 
For a two-dimensional double-well potential and the familiar H\'enon-Heiles 
potential, we obtain very good agreement with exact quantum-mechanical
calculations. 
We also consider the integrable limit of the scenario which corresponds to the 
bifurcation of a torus from an isolated periodic orbit. For the separable 
version of the H\'enon-Heiles system we give an analytical uniform trace 
formula, which also yields the correct harmonic-oscillator SU(2) limit at 
low energies, and obtain excellent agreement with the slightly coarse-grained 
quantum-mechanical density of states.

\section{Introduction}

The goal of the research area called ``quantum chaos'' is to relate the 
quantum-mechanical and classical properties of a classically chaotic system.
For autonomous Hamiltonian systems, the eigenvalue spectrum is, to the
leading orders in $\hbar$, dominated by the periodic orbits of the classical
system. For chaotic systems, the periodic orbits are isolated in phase space 
and contribute individually to the semiclassical spectral density 
\cite{gutz71,gutzbuch}, while in integrable systems the leading contributions 
come from families of degenerate orbits forming rational tori \cite{stma,beta}.
The most general case is that of a system which is neither integrable nor 
ergodic but exhibits a mixed phase space consisting of regular islands 
separated by chaotic domains. The chaotic regions increase through the 
destruction of rational tori when continuous symmetries are broken, and 
through bifurcations of periodic orbits when the energy or another control 
parameter of the system is increased. Explicit semiclassical trace
formulae have been given for various systems with continous symmetries
\cite{stma,bablo,sm77,crli}, for symmetry breaking through the
destruction of rational tori \cite{ozoha,ozobu,crper,toms,hhun}, 
and for isolated bifurcations \cite{si96,ss97,ss98}. 
However more complicated bifurcation scenarios which usually occur in
realistic physical systems and, in particular, bifurcation cascades 
\cite{maodel,main,mbgu,lamp} still constitute one of the most serious 
problems of the semiclassical theory.

Periodic orbits contribute to the semiclassical density of states individually 
only as long as they remain isolated in phase space, i.e., as long as their actions 
differ by large multiples of $\hbar$. Near bifurcations this condition is violated
and the standard remedy is to determine a collective contribution of all periodic 
orbits participating in the bifurcation. In the neighbourhood of a bifurcation, 
this was achieved in \cite{ozoha,ozobu} using the theory of normal forms based 
on the classification \cite{meyer,bruno} of generic bifurcations with 
codimension one (i.e., bifurcations occurring when one control parameter is 
varied). In all these classes, a central orbit of period $n$ is surrounded by 
$m\geq1$ satellite orbits of period $nm$. The corresponding 
generic bifurcations are called isochronous ($m=1$), period-doubling ($m=2$), 
period-tripling ($m=3$), period-quadrupling ($m=4$), etc. The ``local'' uniform
approximations devel\-oped in \cite{ozoha} fail at large distances from the
bifurcations where the orbits become isolated. In \cite{si96,ss97,ss98} 
``global'' uniform approximations were developed, which interpolate between the 
collective contribution of the orbit cluster near a bifurcation and the sum
of individual contributions of the isolated orbits far from it, as correctly
described by the Gutzwiller trace formula \cite{gutz71}. These global uniform
approximations can, with minimal modifications, also be applied to nongeneric
bifurcations in systems with discrete symmetries \cite{then}. Similar global
uniform approximations have also been derived for nongeneric bifurcations of 
codimension two \cite{sc97,sc98}. Even though such bifurcations occur only when 
two control parameters meet the bifurcation conditions simultaneously, they are 
more generally of relevance because they may appear as sequences of generic 
bifurcations when one of the two parameters is fixed and only the other is 
varied \cite{sadov}. In other 
words: when two generic bifurcations lie so close that the orbits do not become 
isolated between them and hence the corresponding generic codimension-one 
global uniform approximations cannot be used, a description using codimension 
two (or higher) becomes necessary. Such a description was first given in 
\cite{sc98} for codimension two along with a classification of the possible 
generic bifurcation sequences according to catastrophe theory.

In the present paper we study a sequence of two successive isochronous 
pitchfork bifurcations of an isolated periodic orbit. This scenario
which occurs in systems with discrete symmetries is not included in the classification 
of codimension-two bifurcations \cite{sc98} so that at present there
exists no semiclassical approach for it. In fact, it may constitute 
the beginning of a bifurcation cascade in which this sequence is repeated 
infinitely often. Such a cascade can form a geometric progression reminiscent 
of the Feigenbaum scenario \cite{feig} (although there the bifurcations are 
generically period doubling), and the new periodic orbits born at the 
bifurcations may exhibit self-similarity properties
\cite{mbgu,lamp}. Bifurcation 
cascades are frequently found in physical systems with discrete 
symmetries and mixed classical dynamics \cite{maodel,main}, so that a
semiclassical approach to those situations would seem to have been
required long ago. Here we develop a uniform approximation of
codimension two for the contribution of 
a pair of pitchfork bifurcations to the semiclassical density of states and 
test it numerically by comparison with exact quantum-mechanical
calculations. The agreement turns out to be very good. 
The degenerate limit, in which the two pitchfork bifurcations coalesce, occurs 
generically in integrable systems: there a whole family of degenerate orbits, 
forming a torus, is born from the central orbit at the bifurcation. For this 
case we can give analytical expressions for our uniform approximation, and 
numerical calculations for a separable system yield an excellent semiclassical 
approximation to the exact quantum-mechanical density of states.

Our paper is organized as follows. In section 2 we present our new uniform 
approximation, whose detailed derivation is given in appendix A. The uniform 
approximation for the bifurcation of a torus from an isolated orbit in the 
separable limit is discussed in section 3, with its detailed derivation given 
in Appendix B. In section 4 we apply our results to a two-dimensional 
double-well potential and to the familiar H\'enon-Heiles system \cite{hhpaper}, 
as well as to its separable version, and compare them to results of exact 
quantum calculations. An alternative derivation of our uniform approximation 
for the separable limit from EBK quantization is given in appendix C.

\section{Uniform approximation in the non-integrable case}

The density of states of an autonomous system with Hamiltonian $H$ is 
given by the trace of the retarded Green function $G(E)$
\begin{equation}
    g(E) \equiv \sum_n \delta \left(E-E_n\right)
    =-\frac{1}{\pi} \; \Im m \; {\rm Tr} G(E)\,,\qquad G(E)=\frac{1}{E+i\,0^+-H}\,.
\label{dos}
\end{equation}
As usual, we split $g(E)$ into a smooth and an oscillating part:
\be
g(E) = {\widetilde g}(E) + \delta g(E)\,.
\ee
The smooth part ${\widetilde g}(E)$, which semiclassically is determined by all 
periodic orbits of the classical system with zero length \cite{bermoun}, may 
either be determined by the (extended) Thomas-Fermi (TF) model \cite{book} or, 
where this is not analytically possible, by a numerical Strutinsky averaging of 
the quantum spectrum \cite{book,strut}. The periodic orbits of finite length
make up the oscillating part $\delta g(E)$.

The semiclassical contribution to $\delta g(E)$ of any region $\Omega$ on a 
Poincar\'e surface of section (PSS) of the phase space is given by \cite{ozobu,si96}
\begin{equation}
 \delta g_{\Omega}\left(E\right)=\frac{1}{2 \pi^2 \hbar^2}\;\Re e\int_{\Omega} \d q' 
 \d p\; \frac{1}{n} \; \frac{\partial \hat{S}}{\partial E} \;
 \left|\frac{\partial^2 \hat{S}}{\partial p \partial q'}\right|^{1/2}
 \exp\left[\frac{i}{\hbar} \hat{S}\left(q',p,E\right)-\frac{i}{\hbar}q'p
 -\frac{i\pi}{2} \nu\right].
\label{omegaintegral}
\end{equation}
Here $q'$ are the final coordinates and $p$ the initial momenta on the PSS 
transverse to a periodic orbit with period $T$ centered in the origin. 
The $n$-th iterate of the Poincar\'e map is given by its generating function 
$\hat{S}\left(q',p,E\right)$, and the usual canonical relations hold:
\begin{equation}
    \frac{\partial \hat{S}}{\partial q'}=p'\,, \hspace{1cm}
    \frac{\partial \hat{S}}{\partial p}=q\,, \hspace{1cm}
    \frac{\partial \hat{S}}{\partial E}=T\,.
\end{equation}
The periodic orbits are the solutions of
\begin{equation}
    \frac{\partial \hat{S}}{\partial q'}=p\,, \hspace{1cm}
    \frac{\partial \hat{S}}{\partial p}=q'\,,
    \label{statpoints}
\end{equation}
which are the stationary points of the phase in \eq{omegaintegral}. If the 
integrals in \eq{omegaintegral} are calculated in the stationary-phase 
approximation, one obtains the individual Gutzwiller contributions of the 
periodic orbits $\xi$ within $\Omega$:
\begin{equation}
    \delta g_{\xi}\left(E\right)={\cal A}_{\xi}\left(E\right)
    \cos\left(\frac{S_{\xi}\left(E\right)}{\hbar}-\frac{\pi}{2} 
    \mu_{\xi}\right),
    \label{gutzcon}
\end{equation}
where for a two-dimensional system the amplitudes have the form
\begin{equation}
    {\cal A}_{\xi}\left(E\right)=\frac{T_{\xi}\left(E\right)}
    {\pi\hbar\, n_{\xi} \sqrt{\left|{\rm Tr\widetilde{M}_{\xi}}-2\right|}}\,.
    \label{gutzwilleramplitudes}
\end{equation}
The quantities $S_{\xi}$, $T_{\xi}$, $n_{\xi}$, $\widetilde{M}_{\xi}$ and 
$\mu_{\xi}$ are the action, period, repetition number, stability 
matrix and Maslov index of the orbit $\xi$, respectively.
The stationary-phase approximation yields good results
only if the periodic orbits are isolated in phase space.
Near a bifurcation this condition is not fulfilled so that one has to 
perform the integrals in \eq{omegaintegral} collectively over the 
whole periodic orbit cluster involved in the bifurcation. To this 
purpose, one inserts the normal form of 
the generating function $\hat{S}\left(q',p,E\right)$ into \eq{omegaintegral}
and solves the resulting integrals exactly.

A sequence of two period-doubling bifurcations of periodic orbits is not generic 
because it would imply a jump in the stability of the central periodic 
orbit \cite{sc98}. On the other hand, an isochronous bifurcation creating a 
new orbit with a degeneracy factor of two is equivalent 
to a generic period-doubling bifurcation \cite{then}. The degeneracy factor 
two has to originate from a two-fold discrete symmetry of the system. Due to 
the behaviour of Tr$\widetilde{M}_{\xi}$ near the bifurcation, a generic
period-doubling bifurcation is often called a pitchfork bifurcation. The case 
of interest here is a sequence of two such nongeneric pitchfork bifurcations which 
can arise successively from the same central periodic orbit in systems with
discrete symmetries such as studied in \cite{mbgu,lamp}. For this scenario we 
propose the new normal form
\begin{eqnarray}
    \hat{S}\left(q',p,E\right)&=&S_0\left(E\right)+q'p
    -\frac{1}{2} \left(\epsilon_1 p^2 +\epsilon_2 q'^2\right)
    -\frac{a}{4} \left(p^2 +q'^2\right)^2 \nonumber \\
    &=& S_0\left(E\right)+q'p
    -\left(\epsilon_1 \cos^2 \phi+\epsilon_2 \sin^2 \phi\right)I
    -aI^2\,.                                                         \label{nf}
\end{eqnarray}
Here $q'=\sqrt{2I} \sin \phi$ and $p=\sqrt{2I} \cos \phi$ define the polar
coordinates $(I,\phi)$ on the PSS, with the central orbit sitting at $I=0$. 
The parameters $\epsilon_1$ and $\epsilon_2$ measure the distance to the 
bifurcations and become zero at the first and second bifurcation, respectively. 
Inserting the angles $\phi=0$ and $\phi=\pi/2$ into \eq{nf}, one obtains the 
respective generic normal forms of the period-doubling bifurcations \cite{ss97} 
corresponding to two cusp catastrophes:
\begin{equation}
    \hat{S}\left(q'\left(0,I\right),p\left(0,I\right),E\right)
    -S_0\left(E\right)-q'\left(0,I\right)p\left(0,I\right)=
    -\frac{\epsilon_1}{2} p^2-\frac{a}{4} p^4
    \label{perioddoubling1}
\end{equation}
and
\begin{equation}
    \hat{S}\left(q'\left(\frac{\pi}{2},I\right),
    p\left(\frac{\pi}{2},I\right),E\right)
    -S_0\left(E\right)-q'\left(\frac{\pi}{2},I\right)
    p\left(\frac{\pi}{2},I\right)=
    -\frac{\epsilon_2}{2} q'^2-\frac{a}{4} q'^4\,.
    \label{perioddoubling2}
\end{equation}
The stationary points of \eq{perioddoubling1} and \eq{perioddoubling2} are the 
stationary points of \eq{nf} as well. The dependence of the topology of \eq{nf} 
on the parameters $\epsilon_i$ is sketched in Fig. \ref{nfcnplts}. 

The period-doubling bifurcations always 
have a real side where the central orbit as well as its satellite orbits are real,
and a complex side where the central orbit is real but the satellite orbits are 
complex ghost orbit. We introduce a parameter $\sigma_i$ which is +1 on the real 
side and -1 on the complex side of the pitchfork bifurcation $i$ with $i=1,2$.
Additionally, the sign of the difference between the actions $S_i$ of the new 
satellite orbits and the action $S_0$ of the central orbit is indicated by
$\tilde{\sigma}_i \equiv {\rm Sign} \left(\Delta S_i\right)$ and 
$\Delta S_i \equiv S_i-S_0$ with $i=1,2$.

\Figurebb{nfcnplts}{55}{560}{483}{625}{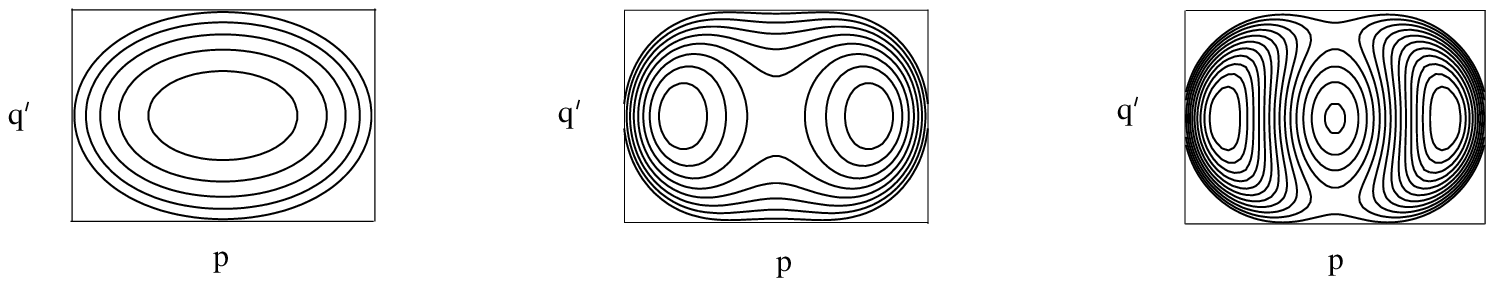}{2.1}{16.6}{
Contour plots of the normal form \eq{nf} in dependence 
of the parameters $\epsilon_i$ for the case $a=-1$. {\it From left to right:} 
$\epsilon_2<\epsilon_1<0$, $\epsilon_2<0<\epsilon_1$ and $0<\epsilon_2<\epsilon_1$.
}

\vspace*{-0.5cm}
The uniform approximation describing the contribution of the orbit cluster 
involved in the bifurcation sequence is derived in detail in appendix A. It reads
\begin{equation}
    \delta g\left(E\right)=\frac{1}{4 \pi^2 \hbar^2} \; \Re e \left\{
    e^{i\,\left(\frac{1}{\hbar} S_0-\frac{\pi}{2} \nu \right)}
    \!\int_0^{2 \pi} d \phi \left[\alpha_0 F_0\left(\phi\right)+
    \alpha_1 F_1\left(\phi\right)+\alpha_2 F_2\left(\phi\right)\right]\right\},
    \label{unapprox}
\end{equation}
where the functions $F_i\left(\phi\right)$ with $i=0,1,2$ are given by
\begin{equation}
    F_0\left(\phi\right)=e^{\frac{i}{\hbar} \frac{\tilde{\sigma}_i}{4}
    \tilde{\epsilon}^2(\phi)} \sqrt{\frac{\pi \hbar}{2}} 
    \left\{\!\frac{1}{\sqrt{2}}\, e^{-i\frac{\pi}{4} \tilde{\sigma}_i}\!
    +\sigma\!\left[ C\!\left(\!\sqrt{\frac{\tilde{\epsilon}^2(\phi)}
    {2 \pi \hbar}}\,\right)\!-i\tilde{\sigma}_i
    S\!\left(\!\sqrt{\frac{\tilde{\epsilon}^2(\phi)}
    {2 \pi \hbar}}\,\right)\right]\!\right\},                    \label{f0}
\end{equation}
\begin{equation}
    F_1\left(\phi\right)=-\frac{1}{2 \tilde{\sigma}_i} 
    \left[i \hbar+\tilde{\epsilon}
    \left(\phi\right) F_0\left(\phi\right)\right], \qquad
    F_2\left(\phi\right)=-\frac{i \hbar}{2 \tilde{\sigma}_i} 
    \left[1-\frac{\tilde{\epsilon}\left(\phi\right)}{2\tilde{\sigma}_i}-
    \frac{\tilde{\epsilon}^2(\phi)}{2i\hbar \tilde{\sigma}_i}\, 
    F_0(\phi)\right],                                          
\end{equation}
and we have used
\begin{equation}
    \epsilon_i=-2 \sigma_i \tilde{\sigma}_i\sqrt{\left|\Delta S_i\right|},
    \hspace{1cm}
    \tilde{\epsilon}\left(\phi\right)=
    \epsilon_1 \cos^2 \phi+\epsilon_2 \sin^2 \phi\,,\hspace{1cm}
    \sigma=-\tilde{\sigma}_i  
    {\rm Sign}\left(\tilde{\epsilon}\left(\phi\right)\right).
    \label{nfcoefficients}
\end{equation}
For the evaluation of \eq{f0} - \eq{nfcoefficients}, any 
of the $\tilde{\sigma}_i$ can be used due to their identical values as described 
in appendix A. The coefficents $\alpha_0$, $\alpha_1$ and $\alpha_2$ are given 
as solutions of the linear system of equations
\begin{equation}
{\cal A}_0=\frac{\alpha_0}{\pi\hbar\sqrt{\left|\epsilon_1\epsilon_2\right|}}\,,
\qquad {\cal A}_1=\frac{\alpha_0-\frac{\alpha_1}{2} \epsilon_1+\frac{\alpha_2}{4} 
\epsilon_1^2}{\pi\hbar\sqrt{\left|-2\epsilon_1\epsilon_2+2\epsilon^2_1\right|}}\,,
\qquad {\cal A}_2=\frac{\alpha_0-\frac{\alpha_1}{2} \epsilon_2+\frac{\alpha_2}{4} 
\epsilon_2^2}{\pi\hbar\sqrt{\left|-2\epsilon_1\epsilon_2+2\epsilon^2_2\right|}}\,,
\end{equation}
where the amplitudes ${\cal A}_i$ with $i=0,1,2$ are given in 
\eq{gutzwilleramplitudes}. 
$C\left(x\right)$ and $S\left(x\right)$ are the standard Fresnel 
functions \cite{abro}. The index $\nu$ is related to the Maslov index $\mu_0$ of 
the central orbit by
\begin{equation}
    \nu=\mu_0-\left(\sigma_1+\sigma_2\right)/2\,.
\end{equation}
All coefficients in \eq{unapprox} are expressed by the quantities 
which appear also in the Gutzwiller contributions \eq{gutzcon}, 
which means that the uniform approximation is invariant under canonical 
transformations.

\section{Uniform approximation for the separable limit}

In the degenerate case $\epsilon \equiv \epsilon_1=\epsilon_2$ the normal 
form \eq{nf} becomes
\begin{eqnarray}
    \hat{S}\left(q',p,E\right)&=&S_0\left(E\right)+q'p
    -\epsilon \left(\frac{p^2+q'^2}{2}\right)
    -a \left(\frac{p^2+q'^2}{2}\right)^{\!\!2} \nonumber \\
&=& S_0\left(E\right)+q'p-\epsilon I-a I^2\,.                \label{intnf}
\end{eqnarray}
Here $\hat{S}\left(q',p,E\right)-q'p$ is independent of the angle $\phi$, which 
means that it refers to an integrable system in which the Hamiltonian depends 
only on the action variables but not on the angles. The normal form \eq{intnf} 
and the corresponding bifurcation scenario has been studied in earlier works 
\cite{ozoha,ozobu,ss98,richens}. What we intend here is to solve the necessary
integrals analytically and express all the coefficients by the actions and the
Gutzwiller or Berry-Tabor amplitudes of the periodic orbits, in order to give 
a final formula which is easy for implementation in actual examples.

The stationary point of the function $\hat{S}-q'p$
now corresponds to a family of periodic orbits, i.e., a rational torus 
which is created from the central orbit at the bifurcation \cite{ozobu}.
It consists of real periodic orbits on one side, whereas its periodic 
orbits have complex coordinates on the opposite side of the bifurcation.
To distinguish between the two sides we introduce a parameter $\sigma$ 
which takes the value $+1$ on the side where the torus is real
and $-1$ on the side where it is complex.

The uniform approximation for this degenerate limit $\epsilon_1=\epsilon_2$ 
of \eq{unapprox} is derived in appendix B and can be given, to the leading 
orders $\hbar$, in analytical form as
\begin{eqnarray}
    \delta g\left(E\right)&=&\frac{{\cal A}_T}{\sqrt{2}}\, \Re e \left\{
    e^{i\,\left(\frac{1}{\hbar} S_T-\frac{\pi}{2} \nu\right)} 
    \left(\frac{1}{\sqrt{2}}\,
    e^{-i\tilde{\sigma} \frac{\pi}{4}}+\sigma \!\left[
    C\left(\!\sqrt{\frac{2\left|\Delta S\right|}{\pi \hbar}}\,\right)-
    i \tilde{\sigma} S\left(\!\sqrt{\frac{2\left|\Delta S\right|}{\pi 
    \hbar}}\,\right)\right]\right)\right\}
    \nonumber \\ \hspace{-2.3cm}
    &+& \sigma \tilde{\sigma}\left({\cal A}_T \sqrt{\frac{\hbar}{4 \pi \left|
    \Delta S\right|}}-{\cal A}_0\right) 
    \cos\left(\frac{S_0}{\hbar}-\frac{\pi}{2} \left(\nu+1\right)\right),
    \label{unapproxintlimit}
\end{eqnarray}
where $S_0$ and $S_T$ are the actions of the central orbit and the torus, 
respectively, and their difference is denoted by
\begin{equation}
    \Delta S \equiv S_T-S_0\,
\end{equation}
with $\tilde{\sigma}\equiv {\rm Sign}\left(\Delta S\right)={\rm Sign}(a)$.
The amplitude ${\cal A}_0$ corresponds to the Gutzwiller amplitude 
\eq{gutzwilleramplitudes} of the central periodic orbit, whereas for the
torus one has to use the Berry-Tabor amplitude ${\cal A}_T$ \cite{beta,richens}. 
The Morse index $\nu$ appearing in \eq{unapproxintlimit} is 
related to the Maslov index $\mu_0$ of the central periodic orbit by
\begin{equation}
    \nu=\mu_0+\sigma \tilde{\sigma}.
\end{equation}

\section{Numerical results}

In order to test the above uniform approximations we apply them to two model 
systems: ($i$) a double-well potential which classically possesses two bifurcation 
cascades, one for approaching the saddle from below and one for approaching it 
from above, ($ii$) the familiar H\'enon-Heiles system as well as its separable 
version. We compare with results from exact quantum calculations and discuss 
the validity of the uniform approximations.

\subsection{A two-dimensional double-well potential}

We study the following Hamiltonian with a double-well potential
\be
H=\frac{1}{2} \left(p^2_x+p^2_y\right)+
  \frac{1}{2} \left(x^2-y^2\right)+\lambda\!\left(y^4-\frac12\,x^2y^2\right)
  +\frac{1}{16\lambda}\,.
\label{hhDW}
\ee

\Figurebb{dwpot}{90}{435}{515}{520}{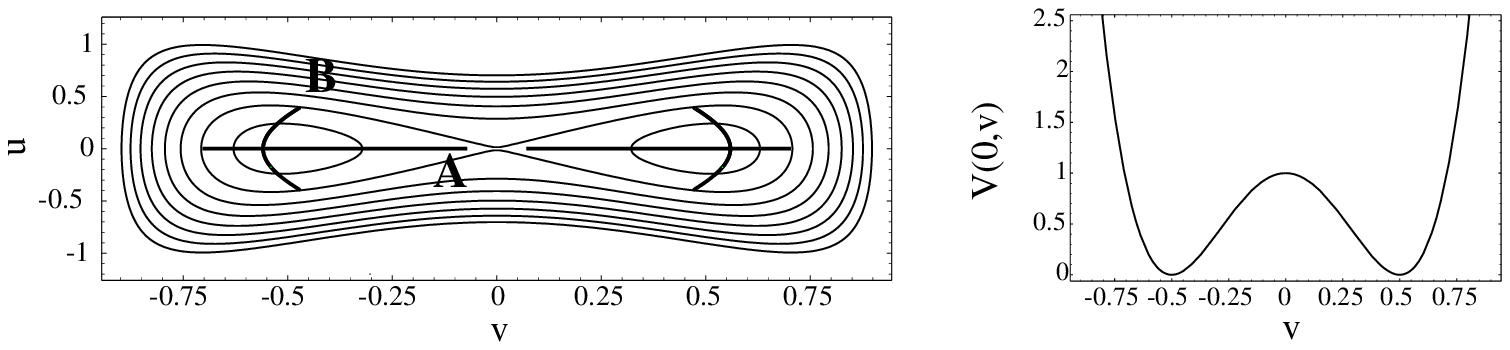}{3.2}{16.6}{
Scaled double-well potential. {\it Left:} contour plot with the 
four shortest periodic orbits $A$ and $B$ evaluated at $e=0.96$. {\it Right:} 
cut of the potential along $u=0$.
}

The potential in \eq{hhDW} has two minima at $x=0$ and $y=\pm 1/2\sqrt{\lambda}$ 
with energy $E=0$, separated by a saddle at $x=y=0$ with energy $E^*=1/16\lambda$.
Using scaled variables $u=\sqrt{\lambda}x$ and $v=\sqrt{\lambda}y$, the classical 
dynamics of the system only depends on one scaled energy variable 
$e=E/E^*=16\lambda E$, with the central saddle at the height $e=1$ (see 
\fig{dwpot}). At a scaled energy $e=9$, the system possesses four other saddles
at $v=\pm 1$ and $u=\pm\sqrt{3}$, over which a particle can escape. At
all energies $e>0$, there exist orbits A and B that librate along and across
the $v$ axis, respectively. Orbit B is stable up to $e=4.778$ and orbit A
undergoes two bifurcation cascades, one approaching the saddle at $e=1$ 
from below, and one approaching $e=1$ from above. We consider here only 
energies $e\leq 1$, for which all periodic orbits appear twice corresponding to 
the two potential wells. In this region, the influence of the continuum above 
$e=9$ can be safely neglected and the quantum spectrum is real and discrete to
a very good approximation. We have obtained it numerically by diagonalisation
of \eq{hhDW} in a finite harmonic-oscillator basis.

The period and action of the A orbit are given analytically in terms of its 
two turning points, for $v>0$ given by
\be
v_1 = \frac12\,\sqrt{1-\sqrt{e}}\,, \qquad v_2 = \frac12\,\sqrt{1+\sqrt{e}}\,.
\qquad (e\leq 1)
\ee
The period becomes
\be
T_A(E) = \frac{\sqrt{2}}{v_2}\,{\bf K}(q),
\ee
and the action is
\be
S_A(E) = \frac{2\sqrt{2}}{3\lambda}\,v_2\left[\frac12\,
         {\bf E}(q)-2\,v_1^2\,{\bf K}(q)\right],
\ee
where ${\bf E}$ and ${\bf K}$ are the complete elliptic integrals \cite{abro}
with modulus $q$:
\be
q=\frac{1}{v_2}\,\sqrt{v_2^2-v_1^2}\,.
\ee
The average (TF) level density of this system (including a factor 2 which 
accounts for the two wells) is given by the integral
\be
g_{TF}(E) = \frac{2\sqrt{2}}{\pi\hbar^2 \sqrt{\lambda}}\int_{v_1}^{v_2}
            \frac{\sqrt{(v_2^2-v^2)(v^2-v_1^2)}}{\sqrt{1-v^2}}\,\d v
\ee
which we could not express in a simple closed form and therefore integrated 
numerically.

At the energy $e=0.91232$ orbit A becomes unstable, creating a stable rotational 
orbit R with Maslov index 5. At $e=0.94272$ orbit A becomes stable again, creating 
an unstable librational orbit L with Maslov index 6. In \fig{dwneworb} the periodic 
orbits R and L are shown together with their complex ``ghost'' predecessors which 
correspond to librations in the real and imaginary parts, respectively. The
bifurcation scenario is seen in the upper left panel of \fig{dwdaten} in terms
of the stability traces.

\Figurebb{dwneworb}{18}{17}{433}{164}{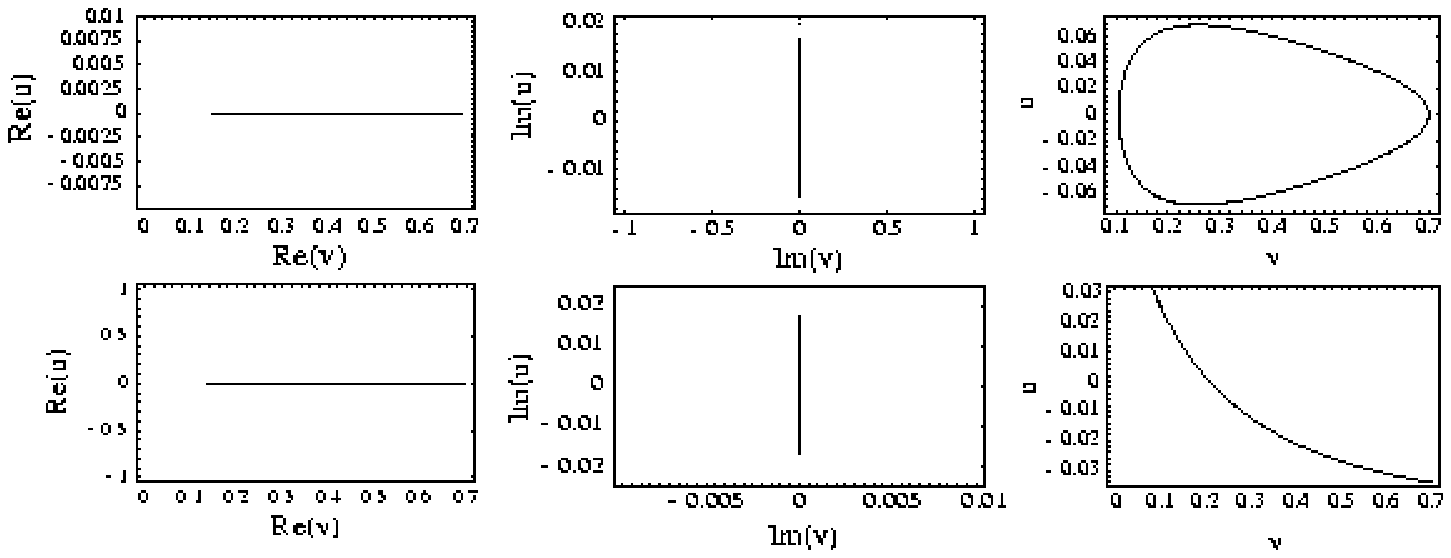}{5.9}{16.6}{
Orbits participating in a pitchfork bifurcation sequence in the double-well
potential \eq{hhDW}. 
{\it Upper row:} Real part (left) and imaginary part (middle) of ghost
orbit R at $e=0.90864$, and real orbit R at $e=0.95$ (right). 
{\it Lower row:} Real part (left) and imaginary part (middle) of ghost
orbit L at $e=0.94$, and real orbit L at $e=0.95$ (right).
}

\vspace*{-0.5cm}
\Figurebb{dwdaten}{140}{375}{550}{715}{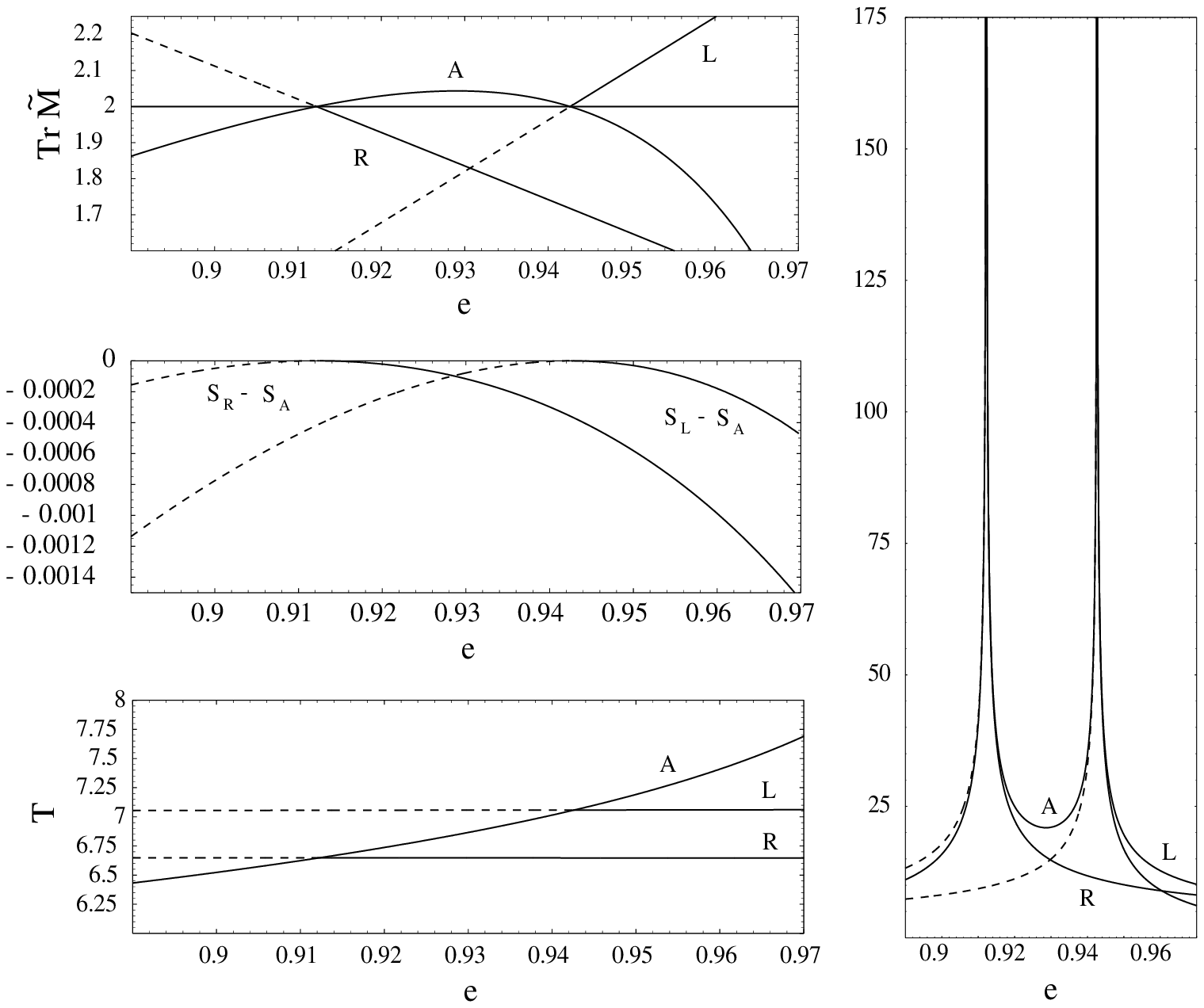}{11.2}{16.6}{
Properties of the periodic orbits A, R and L near their bifurcations in the
double-well potential \eq{hhDW}, plotted versus the scaled energy $e$.
{\it Top left:} stability traces;
{\it middle left:} action differences;
{\it bottom left:} periods; and
{\it right:} Gutzwiller amplitudes (cf. text). The dashed portions of all 
curves correspond to the complex pre-bifurcation ghost orbits.}

\vspace*{-0.5cm}
In \fig{dwdaten} we also show the action differences and periods of the three 
orbits, as well as their Gutzwiller amplitudes, plotted versus the scaled energy 
$e$. As shown analytically in \cite{ss97}, the asymp-

\Figurebb{dwun004}{20}{20}{795}{345}{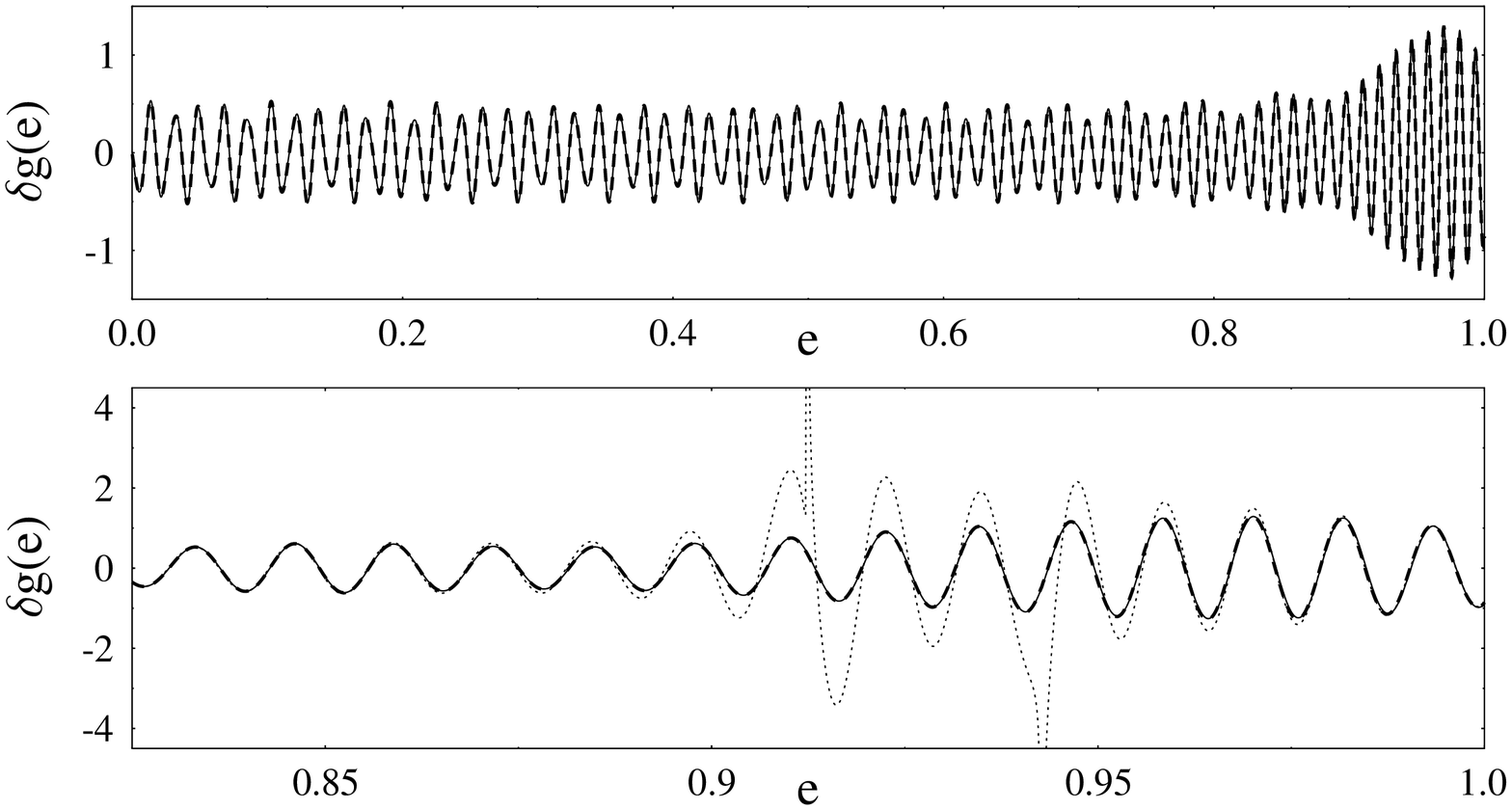}{6}{16.6}{
Oscillating part of density of states in the double-well potential \eq{hhDW}.
{\it Solid line:} exact quantum result obtained with $\lambda=0.0008$. 
{\it Dashed line:} uniform approximation including isolated contribution of 
orbit $B$. {\it Dotted line:} sum of Gutzwiller contributions of isolated orbits, 
diverging at the two lowest bifurcations of the A orbit. (The other bifurcations, 
lying at $e>0.9998$, cannot be seen at this resolution.)
Coarse-graining by Gaussian convolution with energy width $\gamma=0.5$.
}

\vspace*{-0.5cm}
\noindent
totic divergences of the 
amplitudes of the central orbit (here A) and the satellite orbits (here R and L) 
must differ by a factor $\sqrt{2}\,$; this factor has been included in the right 
panel of \fig{dwdaten} in order to confirm this fact numerically.

Using these numerical results we now evaluate the 
uniform approximation \eq{unapprox} for the joint contribution of the orbits
A, R and L. The B orbits are included in the standard Gutzwiller approximation, 
since they stay isolated at all energies and do not interfere with the other 
orbits. In \fig{dwun004} the result is shown together with the result of an exact 
quantum-mechanical diagonalization done for $\lambda=0.0008$. One can recognize 
that the uniform approximation tremendously improves over the diverging standard 
Gutzwiller approximation (dotted line), leading to a perfect agreement with 
quantum mechanics up to the saddle at $e=1$. Here, as well as in all following 
comparisons with quantum mechanics, we have coarse-grained the density of states 
by convolution with a Gaussian over an energy interval $\gamma$. In the 
semiclassical trace formulae this leads \cite{book} to the inclusion of an 
exponential factor $\exp\{-(\gamma T_\xi/2\hbar)^2\}$ in the Gutzwiller amplitude 
${\cal A}_\xi$ of each periodic orbit $\xi$, where $T_\xi$ is its period, in 
regions far enough from the bifurcations for the orbits to be isolated. Note that 
in the regions between the two bifurcations, the Gutzwiller approximation is not
valid, so that our codimension-two uniform approximation is indispensible.

\subsection{The H\'enon-Heiles system}

The system of H\'enon and Heiles is given by the Hamiltonian \cite{hhpaper}
\begin{equation}
    H=\frac{1}{2} \left(p^2_x+p^2_y\right)+
    \frac{1}{2} \left(x^2+y^2\right)+\lambda\left(x^2y
     -\frac13\,y^3\right).    \label{hhH}
\end{equation}
When the scaled variables $u=\lambda x$ and $v=\lambda y$ are introduced, the 
scaled total energy in units of the saddle-point energy $E^*=1/6 \lambda^2$ becomes
\begin{equation}
    e=E/E^*=6\left[\frac{1}{2} \left(\dot{u}^2+\dot{v}^2\right)
    +V\left(u,v\right)\right]
    =3 \left(\dot{u}^2+\dot{v}^2\right)+3\left(u^2+v^2\right)+
    6\,vu^2-2\,v^3.
    \label{shhH}
\end{equation}

In the left part of figure \ref{hhfig} we show the equipotential lines of the 
potential part of \eq{shhH} in the ($u$,$v$)-plane together with the three 
shortest periodic orbits A, B and C, evaluated at the scaled energy $e=1$. 
Along the tree mirror axes (dashed lines) the potential is a cubic parabola as shown 
along $u=0$ in the right part of figure \ref{hhfig}. For an arbitrary energy $e\leq1$, 
the turning points of the A orbit are determined as the solutions of the equation
\begin{equation}
e=3v^2+2v^3
\end{equation}
and given by
\begin{equation}
v_1=1/2-\cos\left(\pi/3-\phi/3\right)\,,\qquad
v_2=1/2-\cos\left(\pi/3+\phi/3\right)\,,\qquad
v_3=1/2-\cos\left(\phi/3\right)\,,
\label{turningpoints}
\end{equation}
with
\begin{equation}
\cos\left(\phi\right)=1-2e\,. \hspace{2cm} \left(e \le 1\right)
\end{equation}
As was shown by H\'enon and Heiles, the classical dynamics is 
quasi-regular up to energies of about $e=2/3$ and then becomes 
increasingly chaotic \cite{hhpaper}. 
The $v$ motion of the A orbit with the scaled energy $e_v$ is given by \cite{lamp} 
\begin{equation}
v_A^{(e_v)}(t)=v_1+\left(v_2-v_1\right) \, {\rm sn}^2 \left(s,q\right)
\label{vA}
\end{equation}
in terms of the Jacobi elliptic function \cite{abro} ${\rm sn}\left(s,q\right)$
which depends on the argument $s$ and the modulus $q$, given by
\begin{equation}
s=t\sqrt{\left(v_3-v_1\right)\!/6}\, \qquad {\rm and} \qquad
q=\sqrt{\frac{v_2-v_1}{v_3-v_1}}\,.                            \label{qmod}
\end{equation}
The turning points $v_i$ have to be evaluated according to the equations
\eq{turningpoints} with $e=e_v$.

\Figurebb{hhfig}{45}{50}{749}{360}{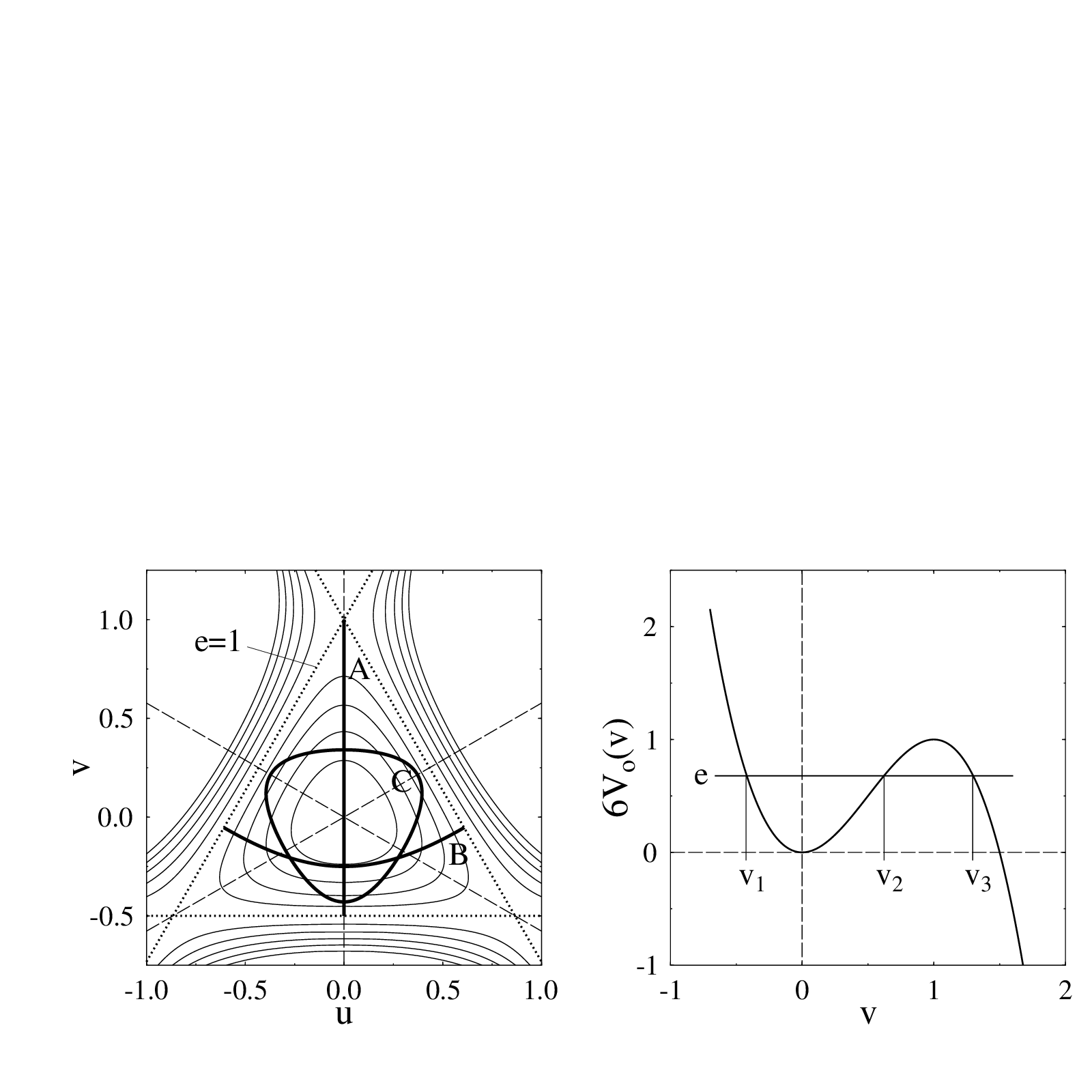}{4.5}{15}{
The H\'enon-Heiles potential. {\it Left:} Equipotential contour lines in
scaled energy units $e$ in the plane of scaled variables $u,v$. 
The dashed lines are the symmetry axes. The three shortest periodic orbits 
A, B, and C (evaluated at the energy $e=1$) are shown by the
heavy solid lines. {\it Right:} Cut of the scaled potential along $u=0$. 
}

\vspace*{-0.5cm}
The periodic orbits of the system have been investigated and classified by 
Churchill \etal~\cite{chur} as well as Davies \etal~\cite{davies}. 
Up to energies of $e \approx 0.97$ there 
exist only three types of periodic orbits with periods $T$ of the order 
of $2 \pi$: the librations A and B, and the rotation C. Due to the 
$D_3$ symmetry of the potential, orbits A and B occur in three 
orientations connected by rotations about $2 \pi/3$ and $4 \pi/3$ in the 
($u$,$v$) plane. Orbit C has a degeneracy of 2 because of the 
time reversal symmetry which corresponds to two different orbits 
with opposite senses of rotation. The orbit B is unstable for all energies and 
the orbit C stays stable for energies below $e=0.8922$ where it becomes unstable 
due to a generic period-doubling bifurcation. The bifurcation cascades of the
A orbit and the orbits generated by them have been studied in detail in 
\cite{mbgu,lamp}; we adapt the names of the orbits given
in these references, whereby the subscripts of the orbit names denote their 
Maslov indices. The A orbit undergoes its first isochronous pitchfork 
bifurcation at an energy $e_1=0.969309$ and the second one at $e_2=0.986709$. 
At the first bifurcation it creates a stable rotational orbit R$_5$ which is 
doubly degenerate due to its two possible senses of rotation. At the second 
bifurcation, it creates an unstable librational orbit L$_6$ which is doubly 
degenerate due to the reflection symmetry of the potential at the $v$ axis.
This scenario repeats itself at higher energies, whereby the pairs of orbits 
R$_7$ and L$_8$, R$_9$ and L$_{10}$, etc, are born. The rotational or 
librational character of these orbits is indicated by the letters R and L, 
respectively. In figure \ref{orbitsborn} the orbits R$_5$ and L$_6$ are 
plotted together with their pre-bifurcation complex ghost orbits. 
All orbits, including A, gain one more degeneracy factor 3 due to the three-fold 
discrete rotational symmetry of the potential, so that the overall degeneracy 
factors of A, R and L orbits are 3, 6 and 6 respectively.

\vspace*{-0.5cm}
\begin{figure}[H]
\begin{center}
\hspace*{0.5cm}
\includegraphics[height=3.0cm]{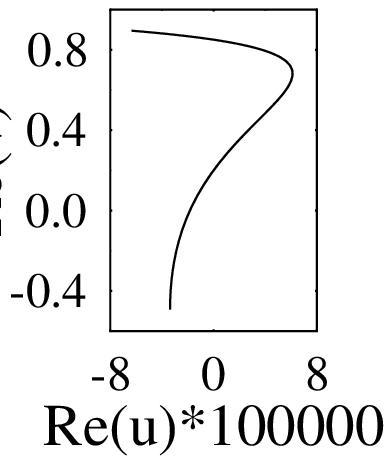}\hspace*{0.2cm}
\includegraphics[height=3.0cm]{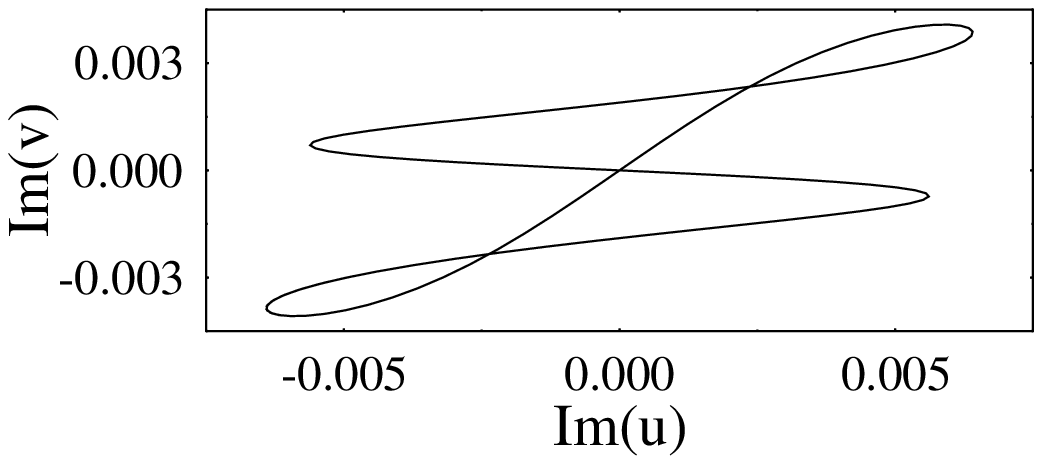}
\includegraphics[height=3.0cm]{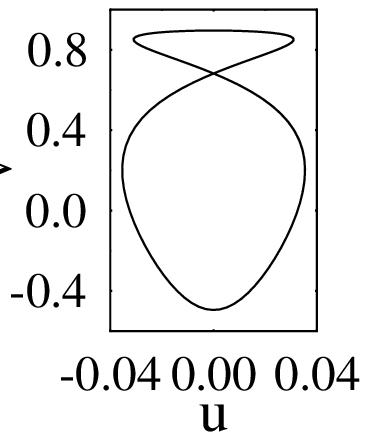}
\\
\vspace*{0.3cm}
\hspace*{0.5cm}
\includegraphics[height=3.0cm]{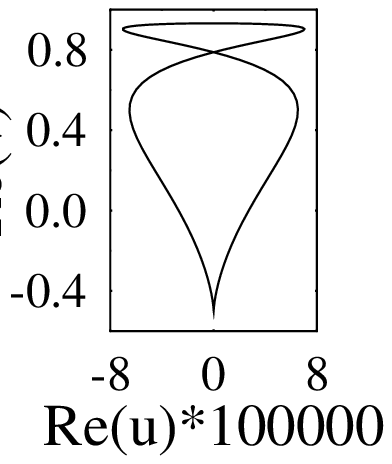}\hspace*{0.2cm}
\includegraphics[height=3.0cm]{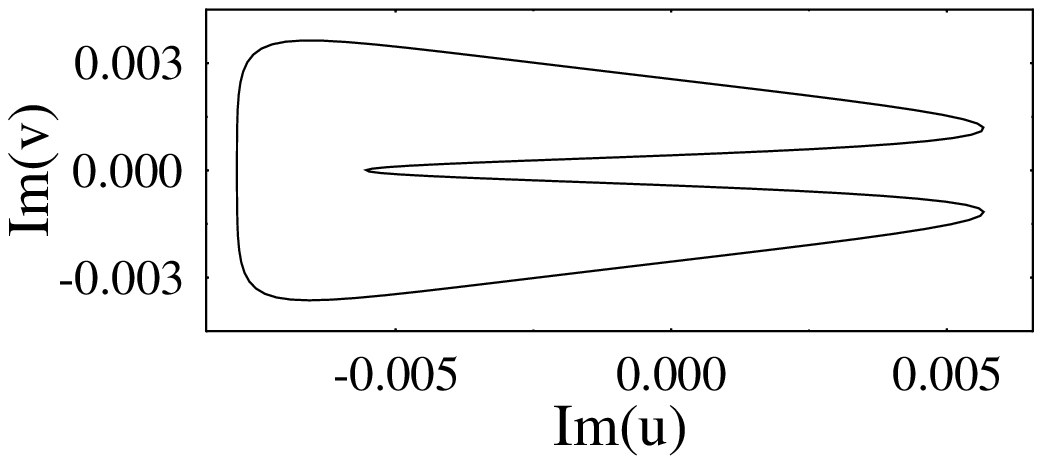}
\includegraphics[height=3.0cm]{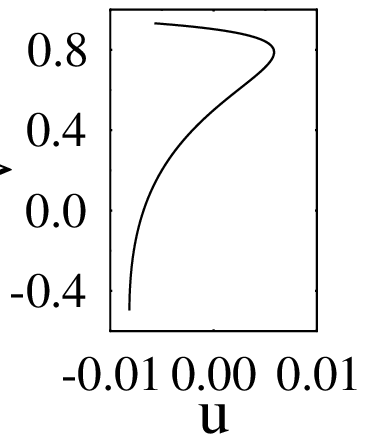}
\parbox{16.6cm}{
\vspace{0.5cm}
\caption[figure]{\renewcommand{\baselinestretch}{0.8}\small\hspace{-0.3truecm}
Orbits born in the first pitchfork bifurcation sequence in the H\'enon-Heiles
potential. 
{\it Upper row:} Real part (left) and imaginary part (middle) of ghost
orbit R$_5$ at $e=0.9690$, and real orbit R$_5$ at $e=0.9798$ (right). 
{\it Lower row:} Real part (left) and imaginary part (middle) of ghost
orbit L$_6$ at $e=0.9864$, and real orbit L$_6$ at $e=0.9870$ (right).}
\label{orbitsborn}}
\end{center}
\end{figure}

\vspace*{-0.7cm}
\Figurebb{pss}{14}{14}{564}{170}{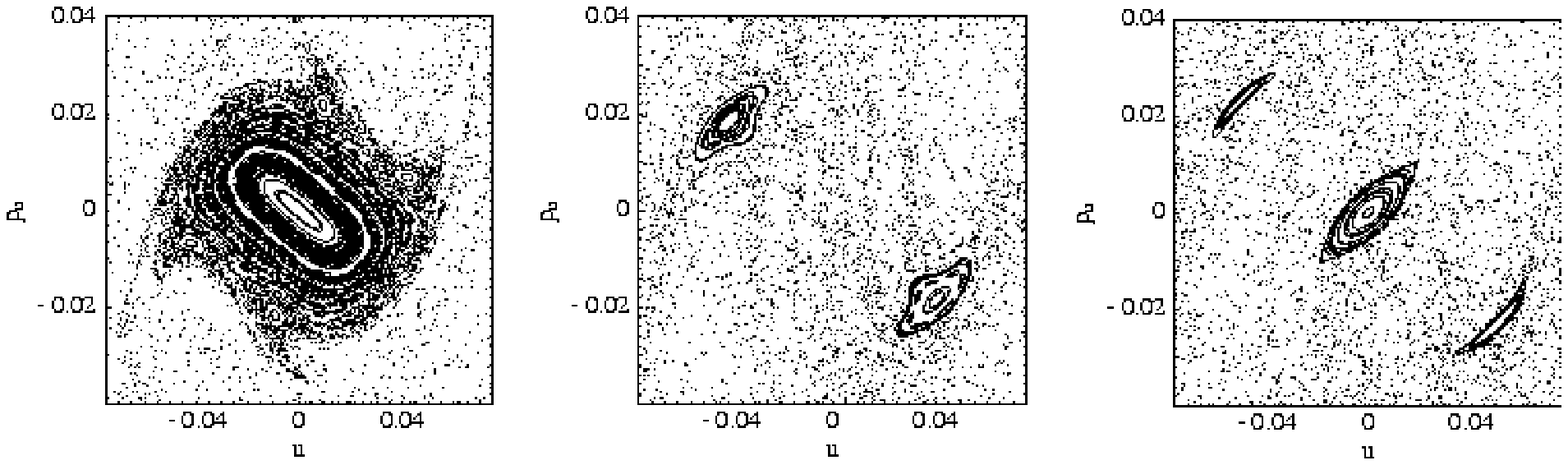}{4.7}{16.6}{
Poincar\'e surfaces of section (PSS) of the scaled H\'enon-Heiles Hamiltonian 
\eq{shhH}, taken for $v=0$. {\it Left:} $e=0.969$; {\it middle:} $e=0.982$; 
{\it right:} $e=0.989$.
}

\vspace*{-0.5cm}
In \fig{pss}, a part of the PSS for $v=0$ is plotted for energies before the 
first pitchfork bifurcation (left), between the first and second bifurcation (middle),
as well as after the second bifurcation (right). The topology in the vicinity of 
the bifurcation sequence is correctly described by the normal form \eq{nf}, as can 
be seen by a comparison with \fig{nfcnplts}.

In evaluating the uniform approximation, one can exploit the fact that the actions
and the periods of the orbit A can be calculated analytically. The action 
is given by
\begin{equation}
S_A\left(E\right)=2\int_{v_1}^{v_2} \sqrt{e-3v^2+2v^3} \; \d v=
\frac{2}{5\lambda^2} \sqrt{6\left(v_3-v_1\right)}
\left[{\bf E}\left(q\right)+c {\bf K}\left(q\right)\right],
\label{SA}
\end{equation}
where the modulus $q$ of the complete elliptic integrals is given in \eq{qmod}. 
The constant $c$ is given by
\begin{equation}
c=-\frac{2}{9} \left(v_3-v_2\right)\left(2v_3-v_2-v_1\right)
\label{q}
\end{equation}
in terms of the turning points $v_i$ ($i=1,2,3$) given in \eq{turningpoints}. 
The period is obtained as
\begin{equation}
T_A\left(E\right)=\frac{\partial S_A\left(E\right)}{\partial E}=2\sqrt{3} 
\int_{v_1}^{v_2} \frac{\d v}{\sqrt{e-3v^2+2v^3}}
=\frac{2 \sqrt{6}}{\sqrt{v_3-v_1}}\, {\bf K}\left(q\right).
\label{TA}
\end{equation}

In \fig{quantities} the quantities needed to evaluate the uniform approximation 
\eq{unapprox} of the density of states are shown as a function of the scaled 
energy $e$. One can see that the stability trace Tr${\widetilde M}_A$ of the
A orbit takes on the values $+2$ at the bifurcation energies. The stability 
traces of the orbits $R_5$ and L$_6$ are also plotted; they 
stay real even for energies $e<e_1$ and $e<e_2$, respectively, where the 
two satellites are complex ghost orbits (which their properties shown by
dashed lines in \fig{quantities}).

In \cite{hhprl}, it was shown that the coarse-grained quantum-mechanical 
density of states of the H\'enon-Heiles potential (obtained with a Gaussian
smoothing width $\gamma=0.25$) can be rather accurately approximated 
semiclassically, using just the isolated orbits A, B and C and their second 
repetitions, for energies far enough from the harmonic-oscillator limit $e=0$. 
In \cite{hhun}, a uniform approximation for the symmetry breaking at $e=0$ was 
developed which continuously interpolates from the harmonic-oscillator limit, 
given in \eq{dgho} below, to the region where the Gutzwiller trace formula for 
the isolated orbits is valid. However, the bifurcations of the A orbit have not 
been treated uniformly in the references \cite{hhun,hhprl}, so that the accuracy 
of the results decreased near the saddle at $e=1$. In \cite{mbgu} the classical 
bifurcation cascade in the H\'enon-Heiles potential was discussed, in which the 
sequence of two successive pitchfork bifurcations repeats itself infinitely often.

\Figurebb{quantities}{105}{346}{556}{630}{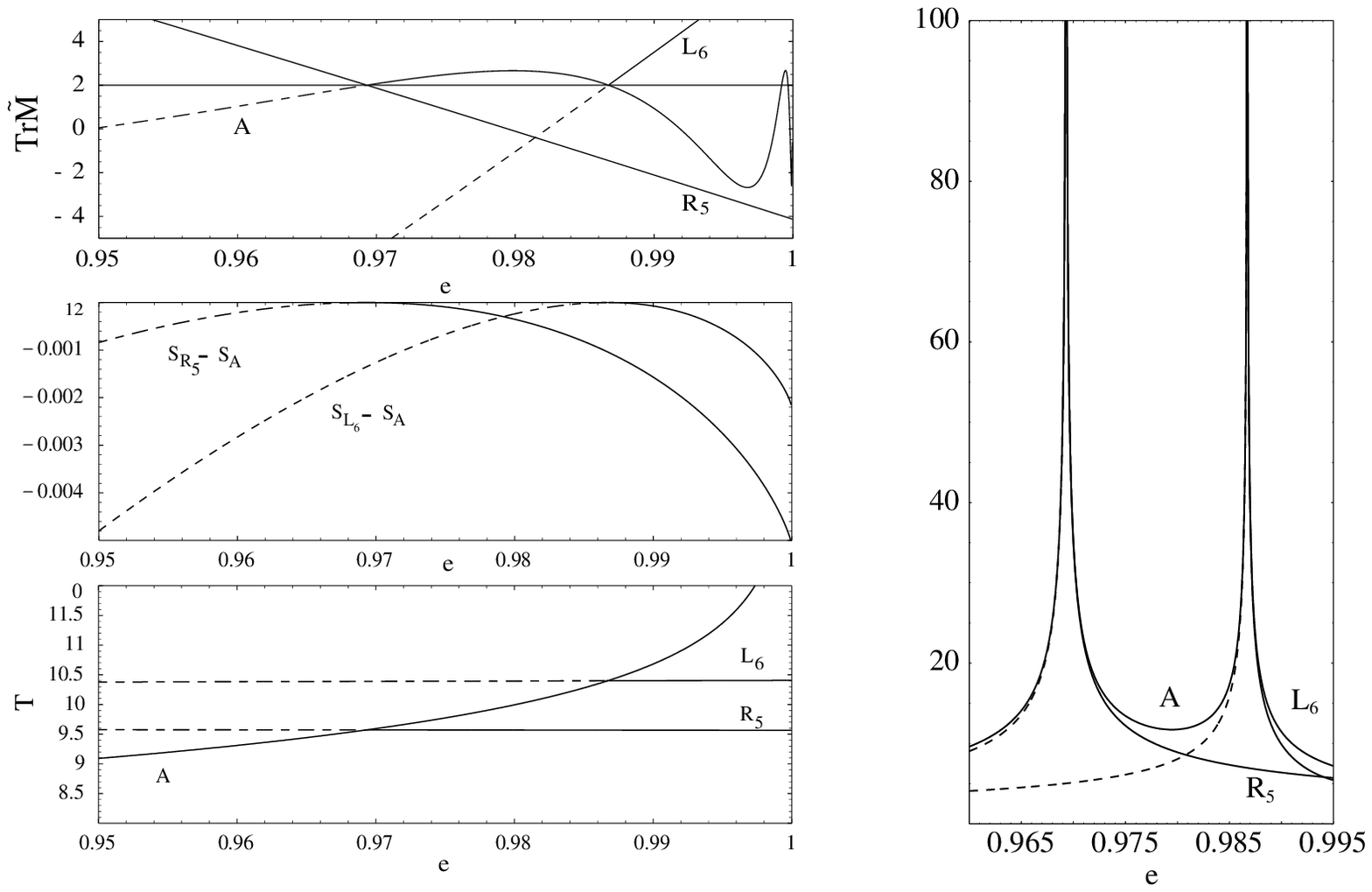}{10.3}{16.6}{
Same as \fig{dwdaten} for the H\'enon-Heiles potential near the first two
bifuractions of the A orbit.
}

Presently we test our uniform approximation \eq{unapprox} to the density of
states against the quantum-mechanical result obtained for $\lambda=0.03$. The 
quantum spectrum was, as in \cite{hhun,hhprl}, obtained by diagonalisation of 
\eq{hhH} in a finite harmonic-oscillator basis -- thus neglecting the effects of 
quantum tunnelling through the barrier. Both quantum and semiclassical results were 
coarse-grained with a Gaussian width of $\gamma=0.4$; for this resolution the 
inclusion of the second repetitions of all periodic orbits in \eq{unapprox} was 
necessary (cf. \cite{hhun}). For the pitchfork bifurcation of the second repetition 
of the orbit C at $e=0.892$, where a double-loop orbit D is created \cite{hhprl},
we used the codimension-one uniform approximation of \cite{ss97}. The upper part 
of \fig{gutzund} shows the entire energy region $0\leq e \leq 1$, whereas the lower 
part shows the zoomed region $0.88\leq e\leq1$. The solid lines give the 
quantum-mechanical result, and the dashed lines the results obtained with our 
uniform approximation \eq{unapprox} for the first two pitchfork bifurcations of the 
A orbit. In the region $e\leq 0.5$, we have included the uniform approximation for 
the symmetry breaking, developed in \cite{hhun}, in order to obtain the correct 
harmonic oscillator limit for $e\rightarrow 0$. The dotted line in the lower
part of the figure corresponds to the sum of the isolated periodic orbits 
according to the standard Gutzwiller trace formula \cite{gutz71}. Here 
the divergences due to the lowest bifurcations of the A and C orbits are clearly 
visible. The uniform result \eq{unapprox}, however, exhibits no divergences and 
its agreement with the quantum result is very satisfactory. The discrepancy
arising at $e\simg 0.992$ can be attributed to the influence of the continuum
that starts at $e=1$ which was not taken properly into account in our quantum
result. In fact, the rather excessive maximum appearing in the latter around 
$e\sim 0.994$ makes us believe that the latter is erroneous, rather than our 
semiclassical result. Note that the uniform approximation properly yields the
asymptotic Gutzwiller result on either side of the double-pitchfork bifurcation.

\Figurebb{gutzund}{40}{40}{795}{340}{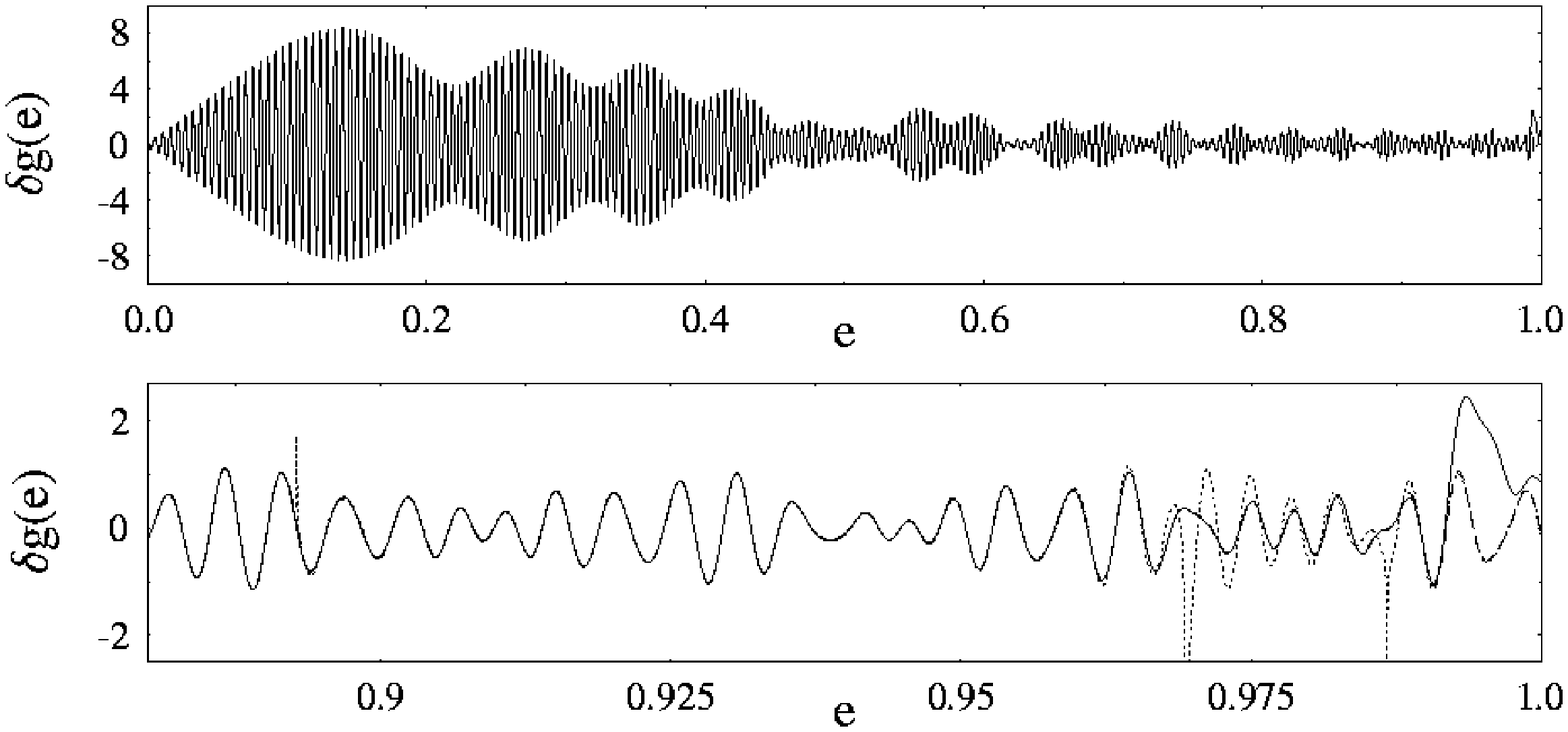}{5.6}{16.6}{
Oscillating part of density of states in the H\'enon-Heiles potential.
{\it Solid lines:} quantum-mechanical results obtained for $\lambda=0.03$. 
{\it Dotted lines:} sum of Gutzwiller contributions \eq{gutzcon} of all 
isolated orbits.
{\it Dashed lines:} codimension-two uniform approximation \eq{unapprox} for 
the orbits A, R$_5$ and L$_6$, including orbits C and D in the codimension-one
uniform approximation of \cite{ss97} and the isolated B orbit.
Coarse-graining with Gaussian width $\gamma=0.4$.
}

\vspace*{-0.5cm}
In the energy region $e>1$ above the barrier, where the spectrum of the 
H\'enon-Heiles Hamiltonian \eq{hhH} is continuous, the oscillating part
of the density of states is determined by the resonances in the continuum. 
In order to test the semiclassical periodic orbit theory in this domain,
it becomes necessary to calculate both the positions and widths of the
resonances. It will then be an interesting question to study which periodic 
orbits are important in the continuum region. Work along these lines is in 
progress \cite{jkpw}. Although the continuum region is also classically 
unbounded, all the R and L orbits bifurcating from the A orbit (which itself 
ceases to exist above $e=1$), as well as the D orbit bifurcating from C,
continue to exist and are bounded at all energies $e>1$ \cite{mbgu,davies}. In
addition, three new orbits librating across the saddles exist in this region
\cite{mbgu,chur}; since they have the shortest periods they are 
expected to play a leading role in the coarse-grained density of states.

\newpage

\subsection{The separable H\'enon-Heiles system}

The H\'enon-Heiles system permits chaotic motion because of the 
nonseparable term $x^2y$ in \eq{hhH}. Omitting this term one obtains 
a system which is separable in $x$ and $y$ and hence integrable:
\begin{equation}
    H=\frac{1}{2} \left(p^2_x+p^2_y\right)+
    \frac{1}{2} \left(x^2+y^2\right)-\frac{\lambda}{3}\, y^3\,.
    \label{sephhH}
\end{equation}
Again using scaled variables $u=\lambda x$ and $v=\lambda y$ the 
scaled energy $e$ in units of the saddle point energy $E^*$ reads
\begin{equation}
    e=E/E^*=6\left[\frac{1}{2} \left(\dot{u}^2+\dot{v}^2\right)
    +V\left(u,v\right)\right]
    =3\left(\dot{u}^2+\dot{v}^2\right)+3\left(u^2+v^2\right)-2\,v^3\,.
    \label{ssephhH}
\end{equation}
Figure \ref{contoursIHH} shows a contour plot of the potential part of 
\eq{ssephhH}
in the $(u,v)$ plane together with the two shortest periodic orbits A 
and B calculated at an energy $e=1$. The two orbits are librations 
along the $u$ and $v$ axes. The potential along the $v$ axis is the same 
as that in the right part of figure \ref{hhfig}, while the potential 
along the $u$ axis is harmonic.
\begin{figure}[H]
\begin{center}
\includegraphics[height=6cm]{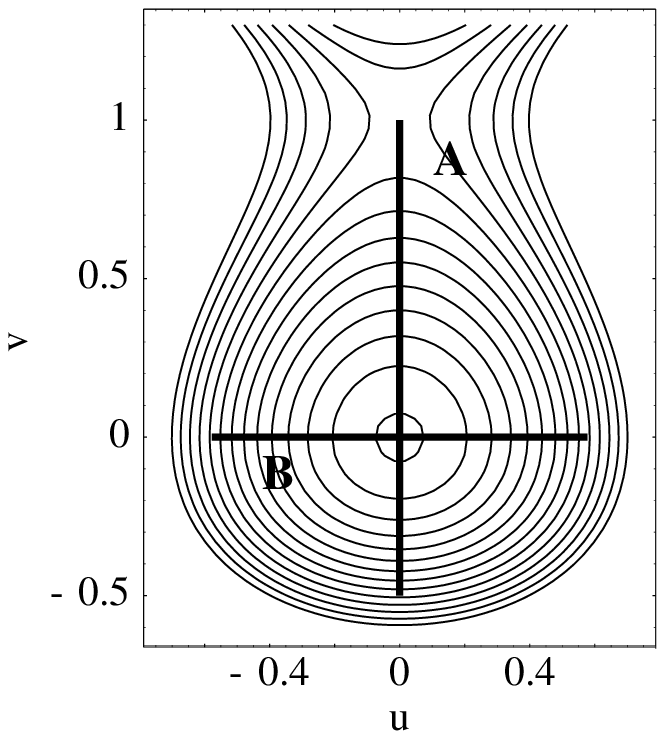}
\parbox{12cm}{
\caption[figure]{\renewcommand{\baselinestretch}{0.8} \small
                                           \hspace{-0.3truecm}
Equipotential lines in the ($u$,$v$) plane for the separable version of the 
H\'enon-Heiles potential. The heavy solid lines show the two shortest periodic 
orbits A and B evaluated at $e=1$.}
\label{contoursIHH}}
\end{center}
\end{figure}

Th actions and periods of the A orbit are given by \eq{SA} and \eq{TA}, respectively.
The trace of its stability matrix is given analytically by \cite{mbgu}
\begin{equation}
{\rm Tr}\widetilde{M}_A(E) = 2 \cos\left(T_A\left(E\right)\right).
\end{equation}
The $u$ motion of the B orbit is harmonic
\begin{equation}
u_B(t)=\sqrt{\frac{e_u}{3}} \sin\left(t+\phi\right)\,,
\end{equation}
where $e_u$ is the conserved scaled energy in the $u$ direction and the phase 
$\phi$ is arbitrary. The action and period of the primitive B orbit are
\begin{equation}
S_B\left(E\right)=2\pi E\,,\qquad
T_B\left(E\right)=2\pi\,.
\end{equation}
The trace of the stability matrix of the B orbit has the constant value 
${\rm Tr}\widetilde{M}_B=+2$, which is consistent with its appearing as
a torus in the asymptotic analysis given in appendix C.

The $k_v$-th repetition of orbit A bifurcates whenever the condition
\begin{equation}
{\rm Tr} \widetilde{M}^{k_v}_A=2 \cos{\left(k_v T_A\right)}=+2  \label{res1}
\end{equation}
is obeyed, which is equivalent to the resonance condition (cf. appendix C.2)
at the bifurcation energies $E_{bif}$
\begin{equation}
k_v T_A\left(E_{bif}\right)=2\pi k_u=k_u T_B\,.                \label{res2}
\end{equation}
Thus, the bifurcations of the A orbit create the rational tori corresponding 
to the $k_v:k_u$ resonances. The new tori form families of degenerate periodic 
orbits that are related by the $U(1)$ symmetry due to the freedom in choosing 
the phase $\phi\in [0,2\pi)$ in their $u$ motion
\begin{equation}
u_T\left(E\right)=\sqrt{\frac{e-e_{bif}}{3}} \, \sin(t+\phi)\,,
\end{equation}
where $e_{bif}$ are the scaled bifurcation energies,
while their $v$ motion is ``frozen'' and identical to that of the A orbit
given in \eq{vA} at the corresponding bifurcation energy:
\begin{equation}
v_T(t)=v_A^{(e_{bif})}(t)\,.
\end{equation}
The actions of the tori become
\begin{equation}
S_T(E)=k_v\,S_A\left(E_{bif}\right)+k_u 2 \pi \left(E-E_{bif}\right),
\end{equation}
so that their periods stay constant at
\begin{equation}
T_T=k_u 2\pi=k_u\,T_B\,.
\end{equation}
Like for all degenerate orbit families, their stability trace is constant:
\be
{\rm Tr}{\widetilde M}_T = +2\,.
\ee

\Figurebb{intdaten}{80}{324}{581}{650}{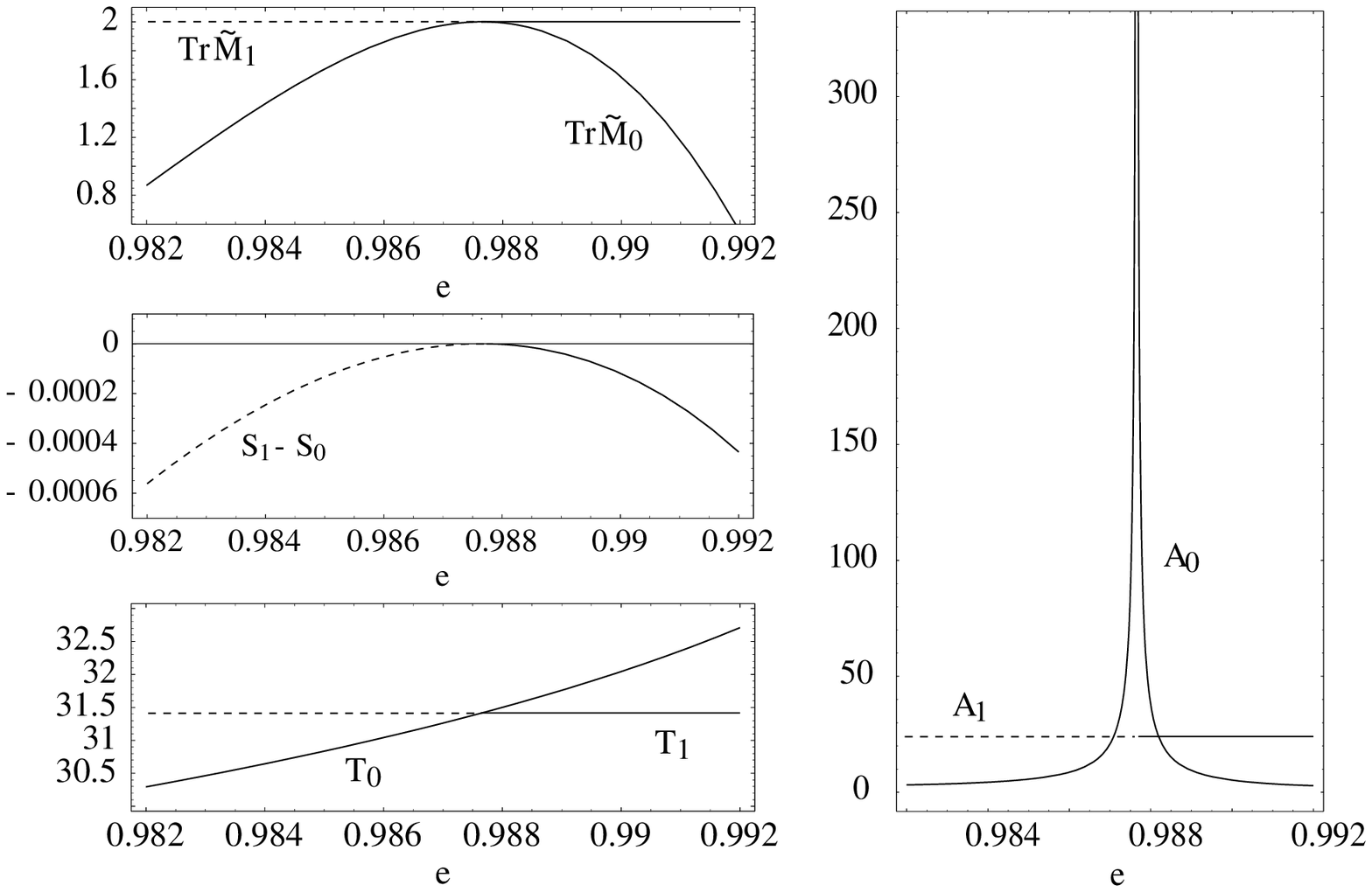}{9.3}{16.6}{
Same as \fig{dwdaten} for the separable H\'enon-Heiles system \eq{sephhH} 
for the bifurcation of the $k_u:k_v=5:3$ resonance at energy $e=0.987655$.
The central A orbit is labeled by ``0'', the bifurcated 5:3 torus by ``1''.
}

We first apply our uniform approximation to the single isolated bifurcation with 
$k_u:k_v=5:3$ which happens at $e=0.987655$. In \fig{intdaten} we show the 
action difference $S_1-S_0=S_T(E)-S_A(E)$, the periods $T_0=3T_A(E)$ and
$T_1=T_T=10\pi$, the traces of the stability matrix, as well as the Gutzwiller 
and Berry-Tabor amplitudes of the isolated A orbit and the 5:3 torus, 
respectively. This figure should be compared with 
\fig{quantities} in which the corresponding quantities are shown for the
non-integrable H\'enon-Heiles potential. Here the two bifurcations
coincide, and instead of the two isolated orbits R$_5$ and L$_6$ created at 
the two bifurcations there, we have here only one torus whose stability trace
has the constant value +2.

These quantities are now used to evaluate the uniform approximation for the
integrable case, given in \eq{unapproxintlimit}. The result is shown in 
\fig{iHHdg} by the dashed line. It is compared to the exact 
quantum-mechanical curve (solid line) obtained for $\lambda=0.04$, as 
well as to the result of including independently the Berry-Tabor contribution 
of the torus and the Gutzwiller contribution of the isolated A orbit which 
diverges at the bifurcation (dotted line). All results have been 
coarse-grained by convolution with a Gaussian with smoothing parameter 
$\gamma=0.1$. We see that the uniform approximation reproduces the quantum 
result very accurately. 

\Figurebb{iHHdg}{30}{40}{795}{265}{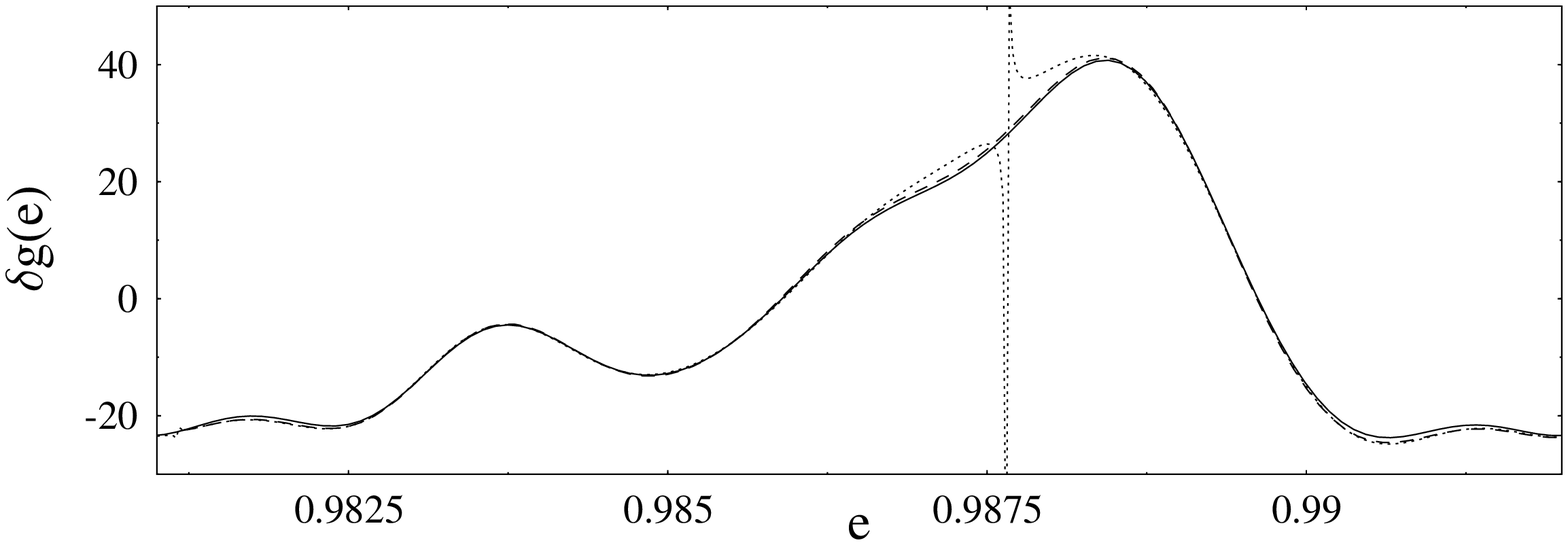}{4.6}{16.6}{
Oscillating part of level density for the separable H\'enon-Heiles system 
\eq{sephhH} near the 5:3 resonance, coarse-grained with a Gaussian width 
$\gamma=0.1$. {\it Solid line:} quantum-mechanical result obtained with 
$\lambda=0.04$; {\it dotted line:} sum of Berry-Tabor 
contribution of 5:3 torus and Gutzwiller contribution of isolated A orbit;
{\it dashed line:} uniform approximation \eq{unapproxintlimit}.
}

So far, we have discussed and tested our uniform approximations for a 
double-pitchfork sequence, based on the normal form \eq{nf}, and its 
separable limit. In appendix C, we give an alternative derivation of the uniform
approximation for the separable limit, starting from the EBK quantization and 
exploiting the convolution property of the density of states for separable systems. 
There we do not require any normal form, but we start from a one-dimensional
integral \eq{dg2gs} for the density of states which by construction is
uniform in the sense that it does not diverge at any energy. By expanding
the amplitude and phase functions of the integrand around the bifurcation
energies $E_{bif}$ up to first and second order, respectively, we arrive
at approximate integrals which precisely correspond to those obtained from
the normal form \eq{nf}, and which can be reexpressed in terms of the
Gutzwiller amplitude of the isolated A orbit and the Berry-Tabor amplitudes
of the rational tori. Furthermore, the starting point \eq{dg2gs} allows us
also to include the limit $e\rightarrow 0$, in which the amplitude of the 
isolated A orbit also diverges, in a uniform way.

Since all amplitudes, actions and periods of the isolated A orbit and 
the tori bifurcating from it can be given analytically for the IHH potential,
it poses no problem to sum over the repetitions of the A orbit and all the
tori bifurcating from them. As shown in detail in the appendix C, this leads 
to the following ``grand'' uniform approximation which is valid and finite
also in the harmonic-oscillator limit $e\rightarrow 0$: 

\newpage

\bea 
\delta g_{uni}(E) \!\!& = & \!\!\!\sum_{k_v=1}^\infty\,\sum_{k_u=k_v}^\infty\!
(-1)^{k_u+k_v}\!\Biggl\{\!\Bigg({\cal A}_{A_{k_uk_v}}(E)-\frac{1}{2}\,\sigma_{k_uk_v}\,
 \sqrt{\frac{\hbar}{\pi\Delta S_{k_u k_v}}}\,{\cal A}_{T_{k_u k_v}}\!\Bigg)
\cos\left[\frac{k_v}{\hbar}\,S_A(E)-\frac{\pi}{2} \right] 
      \nonumber\\ 
&   & \hspace{1.6cm} \;+\;\; \frac{{\cal A}_{T_{k_u k_v}}}{2}\,\Re e \left[\left(
      e^{i\pi/4}[1-\delta_{k_uk_v}]
      +\sqrt{2}\,[C(\xi_{k_u k_v})+iS(\xi_{k_u k_v})]\right)
      e^{\frac{i}{\hbar}S_{T_{k_u k_v}}\!(E)}\right]\!\Bigg\}\nonumber\\
      \cr
& + & \!\!\delta g_{as}^{(A0)}(E) \;+\; \delta g_{as}^{(B0)}(E)\,. \label{dgugra}
\eea
All quantities appearing above are given analytically in equations \eq{tamp}, 
\eq{st}, \eq{aamp} and \eq{sigdel} of the appendix C.

The first term in \eq{dgugra} yields, upon summation over all $k_u$ and $k_v$
and adding the term $\delta g_{as}^{(A0)}(E)$ in the last line, precisely the 
Gutzwiller trace formula \eq{dga} of the isolated A orbit which diverges at the 
bifurcations and at $E=0$. The second term in the first line is a counter term 
from the tori that cancels all divergences of the Gutzwiller amplitudes. 
The second line of \eq{dgugra} yields the Berry-Tabor trace formula 
\eq{dgt} far away from the bifurcations; near the bifurcations it contains the 
Stokes factor that interpolates between the Berry-Tabor amplitudes above and zero 
below the bifurcations, yielding exactly half the Berry-Tabor amplitudes at the 
bifurcations. The two contributions in the last line of \eq{dgugra} are small 
boundary terms, given in \eq{dgb0} and \eq{dga0} of the appendix C, which are 
numerically insignificant but have been included in order to be consistent up to 
order $\hbar^{-1}$ in the amplitudes. 

In the limit $e\rightarrow 0$, where we can neglect all bifurcations, only the 
diagonal terms with $k_u=k_v=k$ contribute. The trace formula \eq{dgugra} then 
leads uniformly to the correct SU(2) harmonic oscillator limit whose trace formula 
is given in equation \eq{ghoiso} of the appendix C:
\be
\delta g_{uni}(E) \quad \longrightarrow 
                  \quad \delta g_{ho}^{iso}(E) =
                  \frac{2E}{\hbar^2}\,\sum_{k=1}^\infty
                  \cos\left(\frac{k}{\hbar}2\pi E\right)
                  \qquad (\hbox{for}\;\; e\rightarrow 0)\,. \label{dgho}
\ee
(The same limit was obtained in a uniform approximation for the full 
non-integrable H\'enon-Heiles potential in \cite{hhun} neglecting, however,
the bifurcations.)

In the figures \ref{dgtest1} and \ref{dgtest2}, we compare the results 
obtained from the grand uniform approximation \eq{dgugra} with those of 
quantum-mechanical calculations for system \eq{ssephhH} with $\lambda=0.04$
(with saddle energy $E^*=104.666$ corresponding to $e=1$),
both coarse-grained by a Gaussian convolution with an energy range
$\gamma=0.1$, including repetition numbers up to $|k_u|,|k_v|\leq 8$ into 
the semiclassical trace formula \eq{dgugra}. Figure \ref{dgtest1} shows
the lowest energy range which exhibits for $e\siml 0.1$ the 
harmonic-oscillator limit \eq{dgho} where the amplitude of $\delta g(e)$ is
linear in $e$. 

\Figurebb{dgtest1}{55}{34}{896}{237}{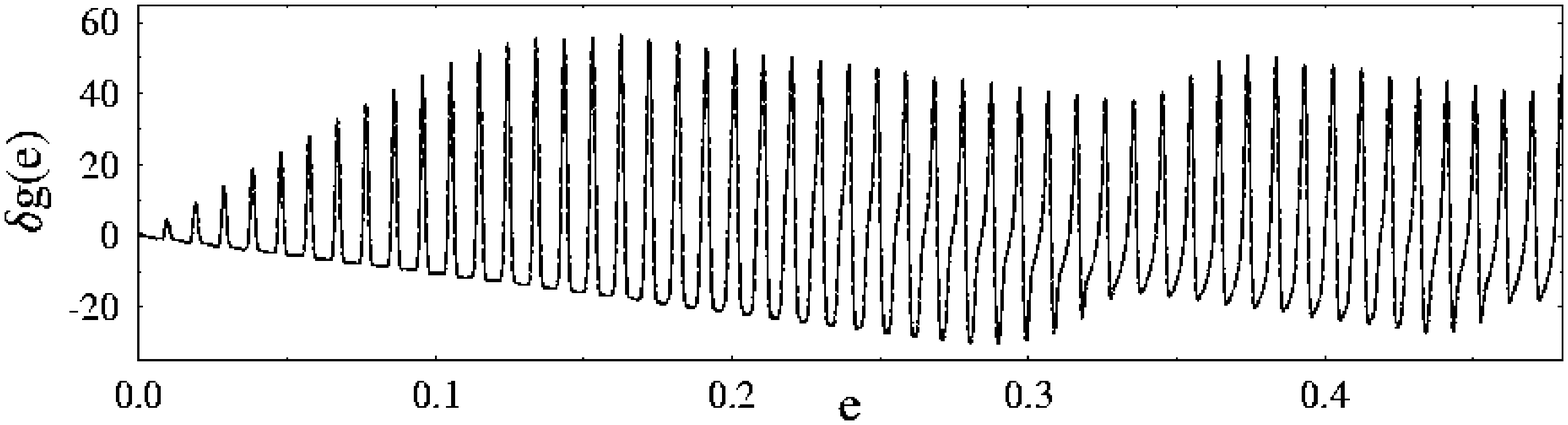}{4.1}{16.6}{
Oscillating part of level density of the separable H\'enon-Heiles system 
\eq{sephhH}, coarse-grained with $\gamma=0.1$. 
{\it Solid lines:} quantum-mechanical result for $\lambda=0.04$. {\it Dashed 
lines:} semiclassical results with $k_u,k_v\leq 8$.
}

\Figurebb{dgtest2}{92}{25}{1010}{640}{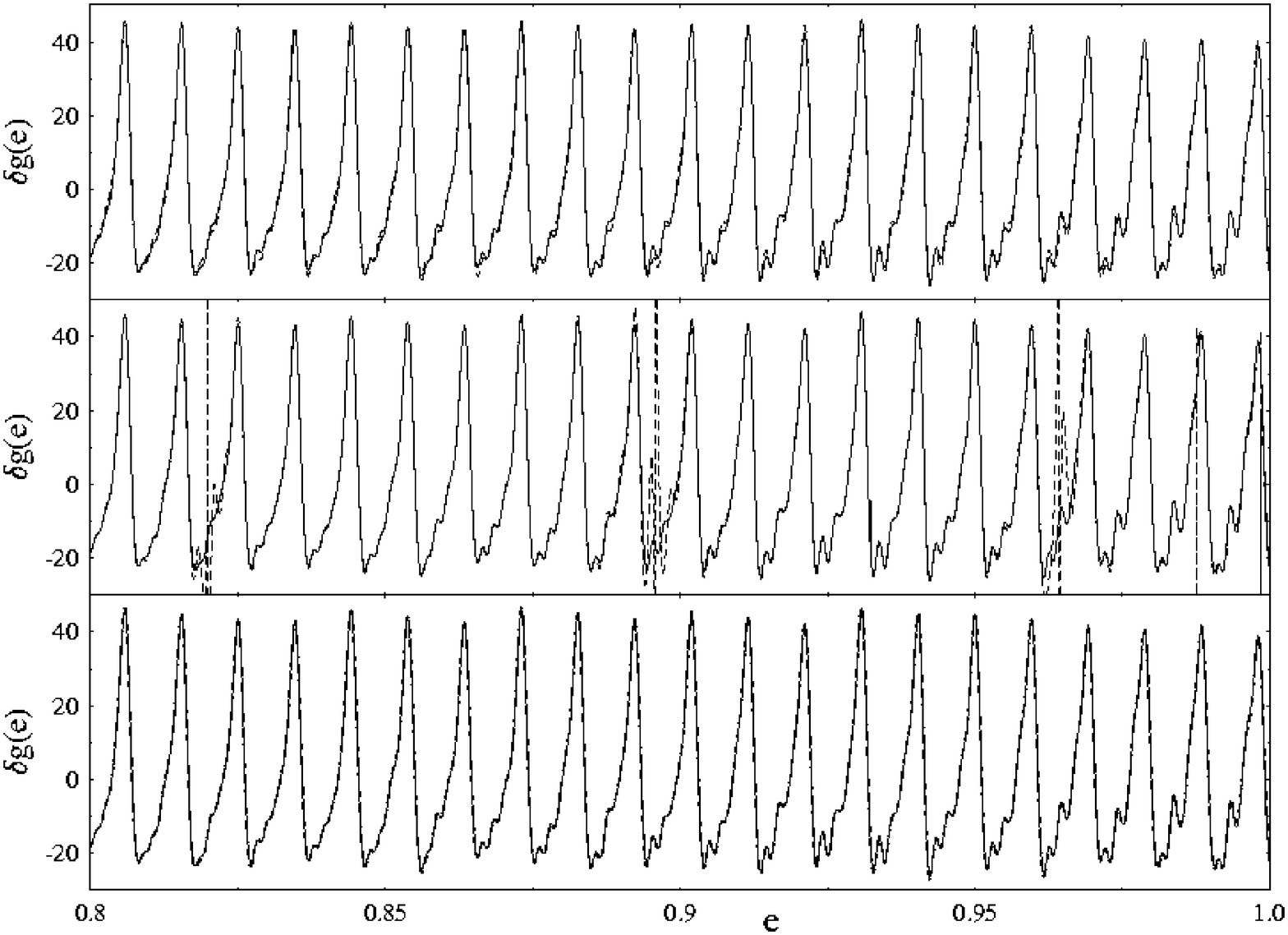}{11.7}{16.6}{
Oscillating part of level density of the separable H\'enon-Heiles system \eq{sephhH}, 
coarse-grained with $\gamma=0.1$. {\it Solid lines:} quantum-mechanical result. 
{\it Dashed lines:} semiclassical results with $k_u,k_v\leq 8$. 
{\it Top:} Berry-Tabor result for the tori. {\it Centre:}
Sum of Berry-Tabor result for the tori plus Gutzwiller result for the isolated
A orbit. {\it Bottom:} uniform approximation \eq{dgugra}.
}

\vspace*{-0.5cm}
In the top panel of \fig{dgtest2} we compare the quantum result to the
standard Berry-Tabor trace formula, given in \eq{dgt} of appendix C, which
takes into account only the tori with semiclassical amplitudes proportional to 
$\hbar^{-3/2}$. In the center panel, we have added to them the A orbit contribution 
described by the Gutzwiller trace formula, given in \eq{dga} of appendix C,
with amplitudes proportional to $\hbar^{-1}$. The latter is seen to diverge at 
all bifurcations corresponding to resonances with $k_u:k_v \geq 5:4$. Between 
the bifurcations, the result is clearly improved by adding the A orbit 
contribution and comes very close to the quantum result. In the bottom panel, 
finally, we show the grand uniform approximation \eq{dgugra} which reproduces 
the quantum result perfectly throughout the whole energy region.
The bifurcation corresponding to the resonances with $k_u:k_v=2:1$ happens at
the scaled energy $e=0.998491$; all bifurcations with $k_u:k_v>2:1$ happen
thus in the top 0.15 percent of the energy scale very near the barrier.
In this region, the bifurcations are lying so densely that their independent
summation in \eq{dgugra} is strictly not justified. However, at the present
resolution of the spectral density this does not appear to affect our 
numerical result. On the other hand, the good agreement which we find in 
\fig{dgtest2} at all lower energies demonstrates that our grand uniform
approximation \eq{dgugra} successfully sums all partial bifurcation cascades 
of the A orbit limited by the repetition numbers $2\leq k_v,k_u\leq 8$.

We should stress that, like for the non-integrable H\'enon-Heiles potential,
the quantum spectrum was obtained here by diagonalisation in a finite
harmonic-oscillator basis. The persistence of our good agreement up to 
$e\simeq 1$ therefore suggests that the barrier tunnelling effects are
negligible -- at least within the resolution given here by the coarse-graining
width $\gamma=0.1$. 

\newpage

\section{Summary, Conclusions and Outlook}

We have derived a codimension-two uniform approximation for the joint contribution 
of the periodic orbits involved in a double pitchfork bifurcation sequence by 
constructing a suitable normal form. This bifurcation scenario only
occurs in systems with discrete 
symmetries and cannot be treated by the codimension-one uniform 
approximations developed in \cite{ss97} due to the vicinity of the two pitchfork 
bifurcations. Furthermore it does not belong to the unfoldings classified in 
\cite{sc98}, so that a new approach became inevitable. We have also studied the 
limit where both pitchfork bifurcations coincide, resulting in the bifurcation 
of a torus from an isolated orbit such as it happens in integrable systems.
For separable potentials, the same uniform approximation could be rederived 
from an EBK trace formula that accounts both for an isolated orbit and for the
tori bifurcating from it.

Our uniform approximation was tested numerically for two well-known systems with mixed 
classical dynamics: a double-well potential and the familiar H\'enon-Heiles system. 
In both cases the uniform approximation was shown to reach the asymptotic
Gutzwiller approximation on either side of the double-pitchfork bifurcation,
while yielding finite amplitudes throughout the whole energy region. The agreement 
of the semiclassical and quantum-mechanical coarse-grained level densities was 
found to be excellent.

Our uniform approximation is only valid as long as the considered pair of 
pitchfork bifurcations is isolated from other bifurcations. In the examples 
studied here, this is the case for the lowest pair of bifurcations of the isolated 
A orbit. Since this orbit undergoes an infinite bifurcation cascade cumulating at 
the barrier energy $e=1$, our approximation will eventually fail for higher 
double-pitchfork bifurcations; the precise energy where this happens depends on 
the value of the nonlinearity parameter $\lambda$. However, for the coarse-grained 
shell structure obtained with a limited resolution (given by a sufficiently large 
Gaussian width $\gamma$), the higher bifurcations become less important and the 
corresponding divergences in the level density cannot be resolved.

For the separable limit of the H\'enon-Heiles system we have obtained analytical
expressions for the uniform approximation. This allowed us to sum over
a large part of the bifurcation cascade corresponding to 
rational $k_u:k_v$ tori with $k_u,k_v\leq8$. The 
resulting grand uniform approximation \eq{dgugra} for the semiclassical density 
of states, which also correctly describes the SU(2) symmetry-restoring limit for 
$e\rightarrow 0$, leads to an excellent agreement with the slightly coarse-grained 
quantum-mechanical density of states even up to the barrier energy $e=1$. 

An extension of the semiclassical analysis of the density of states to the
energy region above the barrier ($e>1$), including a rigorous quantum-mechanical 
determination of the widths and energy shifts due to barrier tunnelling in the
quasi-bound region of the spectrum and of the resonances in the continuum region, 
is in progress \cite{jkpw}.

\section*{Acknowledgments}

The authors want to thank P. Schlagheck, H. Schomerus and T. Bartsch for very 
helpful discussions and S. Fedotkin and A. Magner for stimulating comments. 
We thank A. Jung for assistance with the numerical diagonalisations.
J.K. acknowledges financial 
support from the Deutsche Forschungsgemeinschaft through the Graduiertenkolleg 
GRK 638 "Nonlinearity and Nonequilibrium in Condensed Matter".

\newpage
 
\section*{Appendix A. Derivation of the uniform approximation} 
  
In this section we describe the procedure leading to the uniform approximation
\eq{unapprox} following the ideas outlined in \cite{sc98}.
The semiclassical approximation of the density of states is given by
\begin{equation}
    \delta g\left(E\right) \approx \frac{1}{2 \pi^2 \hbar^2}
    \Re e \int_{\Omega} dq' dp \;
    \Psi\left(q',p\right) \exp\left[\frac{i}{\hbar} 
    \Phi\left(q',p\right)-i\frac{\pi}{2} \nu\right]
    \label{integralcart}
\end{equation}
with a phase function
\begin{equation}
    \Phi\left(q',p\right)=\hat{S}\left(q',p\right)-q'p
\end{equation}
and an amplitude function
\begin{equation}
    \Psi\left(q',p\right)=\frac{1}{n} \frac{\partial \hat{S}}
    {\partial E} \left|\frac{\partial^2 \hat{S}}{\partial q' 
    \partial p}\right|^{1/2}.
    \label{anhngampfkt}
\end{equation}
The integration is done over any region $\Omega$ of the PSS and $\nu$ corresponds to the Morse-Index. We can set $n=2$ in \eq{anhngampfkt} because the nongeneric pitchfork bifurcations are equivalent to generic period-doubling bifurcations \cite{then}.
Using the normal form \eq{nf} of the generating function the phase 
function $\Phi$ can be expressed in canonical polar coordinates $I$ 
and $\phi$ as
\begin{equation}
    \Phi\left(q'\left(\phi,I\right),p\left(\phi,I\right)\right)=
    S_0-\left(\epsilon_1 \cos^2 \phi+\epsilon_2 \sin^2 \phi\right)I
    -a I^2
    \label{phasefunction}
\end{equation}
with
\begin{equation}
    p=\sqrt{2I} \cos \phi\,, \qquad q'= \sqrt{2I} \sin \phi\,.
\end{equation}
The periodic solutions in \eq{statpoints} correspond to the stationary points 
of $\Phi$ at
\begin{equation}
    \frac{\partial \Phi}{\partial \phi}=0\,, \qquad
    \frac{\partial \Phi}{\partial I}=0\,,
\end{equation}
yielding
\begin{equation}
    \sin\left(2 \phi\right)=0 \quad
    \left(\epsilon_1 \cos^2 \phi+\epsilon_2 \sin^2 \phi\right)+
    2aI=0\,.
    \label{stateq}
\end{equation}
There are four solutions of \eq{stateq}: two with $\cos\left(2 
\phi\right)=1$, corresponding to a satellite orbit which is labeled 
by 1 in the following, and two with $\cos\left(2 \phi\right)=-1$ 
corresponding to a satellite orbit labeled by 2. At the stationary points 
the values of $I$ are
\begin{equation}
    I_i=-\frac{\epsilon_i}{2 a}\,,
    \label{statpts}
\end{equation}
where $i=1,2$. With $\sigma_i \equiv {\rm Sign \left(I_i\right)}$ the satellite 
orbit $i$ is real if $\sigma_i=+1$ and represents a ghost solution with complex 
coordinates $q$ and $p$ if $\sigma_i=-1$. The phase function \eq{phasefunction} 
evaluated at the stationary points \eq{statpts} corresponds to the actions of the 
two satellite orbits
\begin{equation}
    S_i=S_0+\frac{\epsilon^2_i}{4 a}\,.
    \label{actions}
\end{equation}
The periods are $T_0=\partial S_0/\partial E$ and 
\begin{equation}
    T_i=T_0+\frac{\epsilon_i}{2 a} \frac{\partial 
    \epsilon_i}{\partial E}\,.
    \label{periods}
\end{equation}
The traces of the stability matrix can be calculated from
\begin{equation}
    {\rm Tr} \widetilde{M}= \left(\frac{\partial^2 \hat{S}}{\partial p \partial
    q'}\right)^{-1} \; \left[1+\left(\frac{\partial^2 \hat{S}}{\partial 
    p \partial q'}\right)^2-\frac{\partial^2 \hat{S}}{\partial p^2} 
    \frac{\partial^2 \hat{S}}{\partial q'^2}\frac{}{}\right],
    \label{formelfuerspur}
\end{equation}
evaluated at the stationary points \cite{si96,ss97,ss98}. One obtains
\begin{equation}
    {\rm Tr} \widetilde{M}_0=2-\epsilon_1 \epsilon_2\,, \hspace{.8cm}
    {\rm Tr} \widetilde{M}_1=2+2\epsilon_1 \epsilon_2-2
    \epsilon^2_1\,, \hspace{.8cm}
    {\rm Tr} \widetilde{M}_2=2+2\epsilon_1 \epsilon_2-2 
    \epsilon^2_2\,. \label{lasttrace}
\end{equation}
For all orbits the actions \eq{actions}, periods \eq{periods} and 
stabilities \eq{lasttrace} are real quantities even 
though the orbits themselves can be complex. This characteristic of
period-doubling bifurcations which is due to a Stokes transition 
was already mentioned in \cite{ss97}.\\
The Maslov indices $\mu_i$ of the periodic orbits are related to the 
Morse index $\nu$ appearing in \eq{integralcart} by
\begin{equation}
    \mu_i=\nu+\frac{1}{2} \left(n_n-n_p\right)\,,
\end{equation}
where $n_n$ and $n_p$ are the number of negative and positive 
eigenvalues of the matrix
\begin{equation}
\Phi''=
\left( \begin{array}{cc}
      \frac{\partial^2 \Phi}{\partial q'^2} & \frac{\partial^2 \Phi}{\partial q'
      \partial p} \\ \frac{\partial^2 \Phi}{\partial q' \partial p} & 
      \frac{\partial^2 \Phi}{\partial p^2} \end{array} \right),
\end{equation}
evaluated at the stationary points.
They follow as
\begin{eqnarray}
    \mu_0&=&\nu+\left({\rm Sign} \left(\epsilon_1\right)+
    {\rm Sign} \left(\epsilon_2\right)\right)/2\,,\\
    \mu_1&=&\nu+\left({\rm Sign}\left(\epsilon_2-\epsilon_1\right)-
    {\rm Sign}\left(\epsilon_1\right)\right)/2\,,\\
    \mu_2&=&\nu+\left({\rm Sign}\left(\epsilon_1-\epsilon_2\right)-
    {\rm Sign}\left(\epsilon_2\right)\right)/2\,.
\end{eqnarray}
For the amplitude function the following ansatz was found to be sufficient:
\begin{equation}
    \Psi\left(\phi,I\right)=\alpha_0+\alpha_1 I+ \alpha_2 I^2.
    \label{ampfunction}
\end{equation}
Equation \eq{integralcart}, expressed in canonical coordinates, now takes 
on the following form
\begin{eqnarray}
    \delta g\left(E\right) &\approx& \frac{1}{4 \pi^2 \hbar^2} \;
    \Re e \; \exp\left[\frac{i}{\hbar} S_0 -i\frac{\pi}{2} \nu \right]
    \int_0^{2\pi} d\phi \; \int_0^{\infty} dI \;
    \left(\alpha_0+\alpha_1 I+\alpha_2 I^2\right) \nonumber \\
    \hspace{-1.5cm}
    &\times& \exp\left\{-\frac{i}{\hbar} 
    \left[\left(\epsilon_1 \cos^2 \phi+\epsilon_2 \sin^2 \phi\right)I
    +aI^2\right]\right\}.
    \label{integralpol}
\end{eqnarray}
The parameters $\epsilon_i$ measure the distance to the bifurcation $i$. They 
are given by the actions \eq{actions} of the new born orbits as 
\begin{equation}
    \epsilon_i=-2 \tilde{\sigma}_i \sigma_i \sqrt{\left|\Delta S_i\right|}\,.
\end{equation}
where we have set
\begin{equation}
    a=\tilde{\sigma}_i\equiv{\rm Sign}\left(\Delta S_i\right).
\end{equation}
In order to achieve a uniform approximation one evaluates \eq{integralpol} in 
stationary-phase approximation at the stationary points which yields
\begin{eqnarray}
    \delta g^{\left(SP\right)}\left(E\right)&=&\frac{1}{\pi \hbar} 
    \frac{\left(\alpha_0+\alpha_1 I_i+\alpha_2 I^2_i\right)}
    {\sqrt{\left|{\rm det} \Phi''\left(I_i\right)\right|}} \nonumber \\
    \hspace{-2cm}
    &\times& \cos \left\{\frac{1}{\hbar} \left[S_0- 
    \left(\epsilon_1 \cos^2 \phi_i+\epsilon_2 \sin^2 \phi_i\right)I_i
    -a {I_i}^2 \right]\right\}.
    \label{gvonE}
\end{eqnarray}
One can now determine the coefficients $\alpha_0$, $\alpha_1$ and 
$\alpha_2$ by identifying
the Gutzwiller Amplitudes ${\cal A}_i$ with
\begin{equation}
    {\cal A}_i=\frac{1}{\pi \hbar}
    \frac{\left(\alpha_0+\alpha_1 I_i+\alpha_2 I^2_i\right)}
    {\sqrt{\left|{\rm det} \Phi''\left(I_i\right)\right|}}\,, \label{amps}
\end{equation}
where $i=1,2,3$. Defining
\begin{eqnarray}
    \hspace{-1cm}
    \tilde{\epsilon}\left(\phi\right)&\equiv&
    \epsilon_1 \cos^2 \phi+\epsilon_2 \sin^2 \phi\,,
    \label{epsilonvonphi}
\end{eqnarray}
the integrals with respect to $I$ in \eq{integralpol} can be calculated analytically 
using
\begin{equation}
    F_n \equiv
    \int_0^{\infty} dI \; I^n e^{-\frac{i}{\hbar} 
    \left[\tilde{\epsilon}(\phi)I+aI^2\right]} =
    \left(i \hbar \frac{\partial}{\partial \tilde{\epsilon}\left(\phi\right)} 
    \right)^n \int_0^{\infty} dI \; e^{-\frac{i}{\hbar} 
    \left[\tilde{\epsilon}(\phi)I+aI^2\right]}.             \label{jn}
\end{equation}
They yield, for $n=0,1$ and 2:
\begin{equation}
    F_0\left(\phi\right)=e^{\frac{i}{\hbar} \frac{\tilde{\sigma}_i}{4} 
    \tilde{\epsilon}^2(\phi)} 
    \sqrt{\frac{\pi \hbar}{2}} 
    \left\{\frac{1}{\sqrt{2}}\, e^{-i 
    \frac{\pi}{4} \tilde{\sigma}_i}+\sigma \!\left[
    C\left(\!\sqrt{\frac{\tilde{\epsilon}^2(\phi)}
    {2 \pi \hbar}}\,\right)- i \tilde{\sigma}_i 
    S\left(\!\sqrt{\frac{\tilde{\epsilon}^2(\phi)}
    {2 \pi \hbar}}\,\right)\right]\right\},
\label{j1}
\end{equation}
\begin{equation}
    F_1\left(\phi\right)=-\frac{1}{2\tilde{\sigma}_i} 
    \left[i \hbar+\tilde{\epsilon}
    \left(\phi\right) F_0\left(\phi\right)\right]\,, \qquad
    F_2\left(\phi\right)=-\frac{i \hbar}{2 \tilde{\sigma}_i} 
    \left[1-\frac{\tilde{\epsilon}\left(\phi\right)}{2\tilde{\sigma}_i}-
    \frac{\tilde{\epsilon}^2(\phi)}{2i\hbar \tilde{\sigma}_i} 
    \,F_0(\phi)\right],
\label{j2undj3}
\end{equation}
with $\sigma \equiv -\tilde{\sigma}_i {\rm Sign}\left(\tilde{\epsilon}
\left(\phi\right)\right)$.
The remaining $\phi$ integral over the interval $\left[0,2\pi\right]$ can 
easily be calculated numerically.

\section*{Appendix B. Derivation of the uniform approximation for the 
separable limit}

In the case $\epsilon_1=\epsilon_2 \equiv \epsilon$ the phase 
function \eq{phasefunction} simplifies to
\begin{equation}
    \Phi\left(q'\left(\phi,I\right),p\left(\phi,I\right)\right)=
    S_0-\epsilon I-a I^2
\end{equation}
and becomes independent of $\phi$, corresponding to an integrable system.
The stationary point of $\Phi$ corresponds to a torus with the radial 
coordinate
\begin{equation}
    I_T=-\frac{\epsilon}{2a}\,.
\end{equation}
With the definition
\begin{equation}
\sigma \equiv {\rm Sign}\left(I_T\right)\,,
\end{equation}
the torus is real if $\sigma=+1$, while it is imaginary if $\sigma=-1$ which can be understood from
\begin{equation}
I=\frac{p^2+{q'}^2}{2}\,.
\end{equation}
The action of the torus becomes
\begin{equation}
    S_T=S_0+\frac{\epsilon^2}{4a}\,,
\end{equation}
and for the period one obtains
\begin{equation}
    T_T=T_0+\frac{\epsilon}{2a} \frac{\partial 
    \epsilon}{\partial E}\,.
\end{equation}
Using \eq{formelfuerspur} one finds that Tr$\widetilde{M}=+2$ which is 
characteristic of an orbit family.
The amplitude function can be derived from \eq{ampfunction}, resulting in
\begin{equation}
\Psi\left(I\right)=\left(\alpha_0+\beta I\right),
\label{ampfunctioniHH}
\end{equation}
with
\begin{equation}
\beta=\left(\alpha_1-\frac{\epsilon}{2a} \alpha_2\right)
\end{equation}
using $\alpha_0$, $\alpha_1$ and $\alpha_2$ from \eq{ampfunction}.
This can be seen by the following integration by parts
\begin{eqnarray}
-\frac{\epsilon}{2a} \int_0^{\infty} dI \; I \exp \left(
\frac{i}{\hbar} \Phi \right)&=&
\frac{1}{2a} \int_0^{\infty} dI \; I \exp \left(\frac{i}{\hbar} \Phi\right)
\left(-\epsilon-2aI+2aI\right) \nonumber \\
&=&
\frac{1}{2a} \int_0^{\infty} dI \; I \exp 
\left(\frac{i}{\hbar} \Phi\right) \left(\frac{\partial \Phi}
{\partial I}+2aI\right) \nonumber \\
&=&\int_0^{\infty} dI \; I^2 \exp \left(\frac{i}{\hbar}
\Phi\right)+{\cal O}(\hbar)\,,
\end{eqnarray}
where the integral that was neglected in the last step is of relative order 
$\hbar$. Thus we obtain, to leading order in $\hbar$, the result 
\eq{ampfunctioniHH}.

The functions in \eq{epsilonvonphi} no longer depend on $\phi$ so that 
the integration over $\phi$ can be performed giving a factor of $2\pi$.
The remaining expression for $\delta g\left(E\right)$ then has the form
\begin{equation}
    \delta g\left(E\right) = \frac{1}{2 \pi \hbar^2} \;
    \Re e \; \exp\left[\frac{i}{\hbar} S_0 -i\frac{\pi}{2} \nu \right]
    \int_0^{\infty} dI \,\left(\alpha_0+\beta I\right)
    \exp\left[-\frac{i}{\hbar} \left(\epsilon I+aI^2\right)\right].
\end{equation}
Exactly the same formula can be found in \cite{ss98} in
relation with a special case of a generic period-quadrupling bifurcation.
Using the integrals \eq{j1}, \eq{j2undj3}, one arrives at
\begin{eqnarray}
    \delta g\left(E\right)&=&
    \frac{1}{\pi \hbar} \frac{\beta}{2a} 
    \cos\left(\frac{S_0}{\hbar}-\frac{\pi}{2} \left(\nu+1\right)\right)
    +\frac{1}{\pi \hbar^{3/2}} \sqrt{\frac{\pi}{2\left|a\right|}}
    \left(\alpha_0-\frac{\beta \epsilon}{2a}\right)
    \nonumber \\ \hspace{-1cm} &\times&
    \Re e \left\{ e^{\frac{i}{\hbar}\left(S_0+
    \frac{\epsilon^2}{4a}\right)-i\frac{\pi}{2} \nu}
    \left(\frac{e^{-i\frac{\pi}{4}\tilde{\sigma}}}
    {\sqrt{2}}+
    \sigma\left[C\left(\sqrt{\frac{\epsilon^2}
    {2\pi \hbar \left|a\right|}}\right)-i \tilde{\sigma}
    S\left(\sqrt{\frac{\epsilon^2}
    {2\pi \hbar \left|a\right|}}\right)\right]\right)\right\}.
    \label{ausgangintnf}
\end{eqnarray}
It remains to express all the parameters by the quantities that enter 
into the asymptotic contributions of the torus and the central orbit. 
A stationary phase approximation of \eq{ausgangintnf} would deliver the 
contribution of the stationary point corresponding to the torus only.
In order to obtain the contribution of the central periodic orbit at 
$I=0$ one has to include also the end-point corrections to the stationary phase 
approximation (cf. appendix C.2). This amounts to an asymptotic expansion of 
the Fresnel functions for large arguments $x \gg 1$ (cf. \cite{abro}). Keeping
their two leading terms
\begin{eqnarray}
    C\left(x\right) &\sim& \frac{1}{2}+\sin\left(\pi x^2/2\right), \\
    S\left(x\right) &\sim& \frac{1}{2}-\cos\left(\pi x^2/2\right),
\end{eqnarray}
leads to
\begin{eqnarray}
    \delta g\left(E\right)&=&\frac{1}{\pi \hbar}  
    \frac{\alpha_0}{\left|\epsilon\right|} \cos\left(\frac{S_0}{\hbar}-
    \frac{\pi}{2}\left[\nu+{\rm Sign} \left(\epsilon\right)
    \right]\right) \nonumber \\ &+& 
    \frac{1}{\pi \hbar^{3/2}} 
    \frac{1+\sigma}{2} \sqrt{\frac{\pi}{\left|a\right|}} 
    \left(\alpha_0-
    \frac{\beta \epsilon}{2a}\right) \cos\left[\frac{S_0
    +\frac{\epsilon^2}{4a}}{\hbar}
    -\frac{\pi}{2} \left(\nu+\frac{\tilde{\sigma}}{2} \right) \right].
\end{eqnarray}
Two asymptotic contributions can be recognized. One is of 
the order $\hbar^{-1}$, corresponding to the central periodic orbit, and 
one is of the order $\hbar^{-3/2}$ which is the torus contribution.
Asymptotically one obtains a torus contribution
only on the real side of the bifurcation $\sigma=+1$ whereas on the
complex side $\sigma=-1$ the torus contribution asymptotically vanishes
even though the torus amplitude itself must not necessarily go to zero.
The fact that it still gives no contribution asymptotically is due to
a Stokes transition of the torus.

Expressing now \eq{ausgangintnf} with the Gutzwiller amplitude ${\cal A}_A$ of 
the central orbit
\begin{equation}
    {\cal A}_0={\cal A}_A=\frac{1}{\pi \hbar} \frac{\alpha_0}{\left|\epsilon\right|}
    \label{aa}
\end{equation}
and the Berry-Tabor amplitude of the torus
\begin{equation}
    {\cal A}_T=\frac{1}{\pi \hbar^{3/2}} \sqrt{\frac{\pi}{\left|a\right|}} 
    \left(\alpha_0-\frac{\beta \epsilon}{2a}\right),
    \label{at}
\end{equation}
and setting $\Delta S \equiv S_T-S_0$ as well as $\nu=\mu_0-
{\rm Sign}\left(\epsilon\right)$ finally yields the uniform approximation 
\eq{unapproxintlimit}.

\section*{Appendix C. Alternative derivation of the uniform approximation 
for the separable limit from EBK quantization}

In this appendix we give an alternative derivation of the uniform
approximation for the separable limit, starting from EBK (or WKB) 
quantization in one dimension and using the fact that the two-dimensional
density of states can be obtained by a convolution of the two one-dimensional
densities of state. We present the general formulae, which to our knowledge
have not been given before in the literature, in the first subsection and
derive from it the known exact trace formula for harmonic oscillators.
In the second subsection we specialize to the integrable H\'enon-Heiles
system and present a uniform trace formula which sums over all bifurcations
and leads to the correct harmonic-oscillator limit for $e\rightarrow 0$.

\subsection*{1. Semiclassical trace formula for separable Hamiltonians}

For a separable Hamiltonian in two dimensions $(u,v)$
\be
H = H_u(u,p_u) + H_v(v,p_v) = E_u + E_v                      \label{hsep}
\ee
the Schr\"odinger equation separates
\bea
{\hat H}\,\Phi_{nm}(u,v) & = & E_{nm}\Phi_{nm}(u,v)\,,\nonumber\\
          \Phi_{nm}(u,v) & = & \phi_n(u)\,\psi_m(v)\,,
     \qquad E_{nm} = \epsilon_n+\varepsilon_m                \label{spec}
\eea
and the exact quantum density of states can be written as a convolution integral
over the two level densities of the one-dimensional systems:
\be
g(E) = \int_0^E g_u(E-E')\, g_v(E')\,\d E'\,,                \label{fold}
\ee
where
\be
g_u(E) = \sum_n \delta(E-\epsilon_n)\,,\qquad
g_v(E) = \sum_m \delta(E-\varepsilon_m)\,,
\ee
and we have assumed $\varepsilon_m,\epsilon_n>0$. 
We now use EBK quantization $(i=u,v;\;n_u=n,\;n_v=m)$:
\be
H_i = H(I_i)\,,\qquad I_i = \frac{1}{2\pi}\,S_i
    = \frac{1}{2\pi}\oint p_i\, \d q_i = \hbar\,(n_i+1/2)\,. 
\ee
and Poisson summation (cf. \cite{beta,book}) to obtain the following 
semiclassical trace formula for each of the one-dimensional level densities
\be
g_i(E) = \frac{T_i(E)}{2\pi\hbar}\sum_{k_i=-\infty}^\infty
         (-1)^{k_i}\cos\left[\frac{k_i}{\hbar}S_i(E)\right],
\ee
which is identical to the Gutzwiller trace formula \cite{gutz71} for a 
one-dimensional system and yields the corresponding EBK (WKB) spectrum. Using 
\eq{fold}, we thus get the two-dimensional trace formula for the separable 
Hamiltonian \eq{hsep}
\bea
\hspace{-1cm}
g(E) & = &\; \frac{1}{(2\pi\hbar)^2}\!\!\sum_{k_u,k_v=-\infty}^\infty\!\!\!
       (-1)^{k_u+k_v}\!\!\int_0^E
       T_u(E-E')\,T_v(E')\cos\!\left[\frac{k_u}{\hbar}S_u(E-E')\right]\nonumber\\
     &   & \hspace*{7.0cm} 
           \times\cos\!\left[\frac{k_v}{\hbar}S_v(E')\right]\!\d E'\,. 
\label{gfold}
\eea
If the convolution integral is done exactly, this trace formula yields the 
spectrum $E_{nm}$ in \eq{spec} in the EBK approximation. The contribution 
from $k_u=k_v=0$ yields the average Thomas-Fermi (TF) level density which 
becomes a simple convolution integral over the primitive periods of the 
two one-dimensional motions:
\be
{\widetilde g}(E) = g_{TF}(E) = \frac{1}{(2\pi\hbar)^2}
                                 \int_0^E  T_u(E-E')\,T_v(E')\,\d E'\,.
\label{gtf}
\ee

The semiclassical trace formula \eq{gfold}, which contains the smooth part 
\eq{gtf}, requires only the classical periods $T_i(E)$ and actions $S_i(E)$
of the one-dimensional systems as an input. Nevertheless, it contains all 
information about the periodic orbits of the two-dimensional system -- not 
only the degenerate families forming two-dimensional rational tori, but also 
the existing isolated orbits as will be shown explicitly in the following. 
It also handles all possible bifurcations uniformly. The formula \eq{gfold} 
therefore goes far beyond the standard trace formulae \cite{beta,crli} for 
integrable systems which only take the leading rational tori into account 
and cannot account for bifurcations. 

The integral in \eq{gfold} can, in general, not be done analytically. For a 
harmonic oscillator
\be
H = \frac12\,(p_u^2+p_v^2)+\frac12\,(\omega_u^2\, u^2+\omega_v^2\, v^2)\,,  
\label{iho}
\ee
we have $S_i(E)=2\pi E/\omega_i$ and $T_i(E)=2\pi/\omega_i$ ($i=u,v$). The 
integral then is elementary and yields
\bea
\hspace{-1.2cm}
g_{ho}(E) & = & \frac{1}{2\pi\hbar\omega_u\omega_v}\!
                \sum_{k_u,k_v=-\infty}^\infty \!\!\!(-1)^{k_u+k_v}\,e^{-i\pi/2}
                \left\{\frac{1}{(k_u/\omega_u-k_v/\omega_v)}\,
                e^{ik_u2\pi E/\hbar\omega_u} \right.\nonumber\\
          &   & \hspace{5.6cm} + \left. 
                \frac{1}{(k_v/\omega_v-k_u/\omega_u)}\,
                e^{-ik_v2\pi E/\hbar\omega_v}\right\}.
\label{dghosum}
\eea
For irrational frequency ratios $\omega_u:\omega_v$, no singularities arise
and using the identity \cite{abro}
\be
\frac{1}{\sin(z)} = \sum_{k=-\infty}^\infty \frac{(-1)^k}{(z-k\pi)}\,,
                   \qquad \qquad (z\neq n\pi)            \label{polesum}
\ee
we can sum the first term in \eq{dghosum} over all $k_v$ and the second term over 
all $k_u$. The result is the exact Gutzwiller trace formula for the irrational 
harmonic oscillator \cite{book,brja} which yields its correct quantum-mechanical 
spectrum
\bea
g_{ho}(E)&=&\frac{E}{\hom_u\hom_v}
          + \frac{1}{\hom_u} \sum_{k_u=1}^\infty \; \frac{(-1)^{k_u}}{
            \sin \left( k_u \pi \frac{\omega_v}{\omega_u} \right)}\,
            \sin \left( k_u \frac{2 \pi E}{\hom_u} \right) \nonumber\\
      & &   \hspace{1.5cm} +\;  
            \frac{1}{\hom_v} \sum_{k_v=1}^\infty \; \frac{(-1)^{k_v}}{
            \sin \left( k_v \pi \frac{\omega_u}{\omega_v} \right)}\,
            \sin \left( k_v \frac{2 \pi E}{\hom_v} \right). 
\label{gho}
\eea
The last two terms contain the sums over the (only) isolated periodic orbits along 
the $u$ and $v$ axes. Corresponding trace formulae for rational frequency ratios 
can be obtained from the above by taking suitable limits \cite{brja}. E.g., in the 
isotropic limit $\omega_u=\omega_v=\omega$ one obtains
\be
g_{ho}^{(iso)}(E) = \frac{E}{(\hom)^2}\left\{ 1 + 2\sum_{k=1}^\infty
            \cos\left(k\frac{2\pi E}{\hom}\right)\right\},      \label{ghoiso}
\ee
which is again a quantum-mechanically exact trace formula in terms of the two-fold 
degenerate families of periodic orbits with SU(2) symmetry, having the primitive 
actions $S(E)=2\pi E/\omega$. Note that the standard methods to derive the trace 
formula for integrable systems \cite{beta,crli} cannot be used for harmonic 
oscillators, since the (``curvature'') tensor of second derivatives of the 
Hamiltonian \eq{iho} with respect to the torus actions $I_i=E_i/\omega_i$ is 
identically zero.

For systems in which the actions $S_i(E)$ are no simple functions, the integral
in \eq{gfold} can in general only be done numerically. This becomes practically 
impossible if one wants to generate the semiclassical EBK spectrum by summing 
\eq{gfold} over all $k_u$ and $k_v$. In the example treated in the next
subsection, we show how an asymptotic evaluation of the integral
can be used to establish the relation to the Berry-Tabor type trace formula
for the tori and the Gutzwiller trace formula for isolated A orbit, and to 
derive the same uniform approximation for the bifurcations of the isolated A
orbit as we have obtained in appendix B using the normal form theory.

\subsection*{2. Asymptotic evaluation for the separable H\'enon-Heiles system 
                and global uniform approximation}

In the following we specialize to the integrable H\'enon-Heiles (IHH) system 
\eq{sephhH}, expressed in the scaled coordinates $u,v$ as in \eq{ssephhH}. 
Here $T_u=T_B=2\pi$, $S_u(E)=S_B(E)=2\pi E$, and the period $T_A(E)$ and action 
$S_A(E)$ of the $v$ motion are given in equations (\ref{SA}) -- (\ref{TA}). 
The TF level density is then given by
\be
g_{TF}(E) = \frac{1}{2\pi\hbar^2}\,S_A(E)             \label{gtfhhi}
\ee
and the oscillating part can be written as
\be
\hspace{-1.2cm}
\delta g(E) = \frac{1}{2\pi\hbar^2}
              {\sum_{k_u,k_v=-\infty}^\infty}{\!\!\!\!\!\!'}\;
              (-1)^{k_u+k_v}\!\int_0^E T_A(E_v)\,e^{i\left[k_vS_A(E_v)
              +2\pi k_u(E-E_v)\right]/\hbar}\,\d E_v\,,    \label{dg2gs}
\ee
where the prime indicates that the TF contribution from $k_u=k_v=0$ must
be left out. Note that upon independent summations over $k_u$
and $k_v$ the imaginary parts cancel, consistently with the general real
expression \eq{gfold}.

In order to evaluate the integral in \eq{dg2gs} in the semiclassical limit
$\hbar\rightarrow 0$, we use the following general formula \cite{erde,wong}
\bea
\hspace{-1.5cm} \int_a^b T(x)\, e^{iS(x)/\hbar}\, \d x         
 & \sim & \sum_i T(x_i)\,\sqrt{\frac{2\pi\hbar}{|S''(x_i)|}}\,
           e^{iS(x_i)/\hbar+i\,{\rm Sign}[S''(x_i)]\,\pi/4} \nonumber\\
 & & \;\;  + T(b)\,\frac{\hbar}{S'(b)}\,e^{iS(b)/\hbar-i\pi/2}\,
           + \; T(a)\,\frac{\hbar}{S'(a)}\,e^{iS(a)/\hbar+i\pi/2}, \label{wongf}
\eea
neglecting corrections of higher order in $\hbar$. Equation \eq{wongf}
is a generalization of the standard stationary-phase integration,
taking into account the end-point corrections whose contributions are
of order $\hbar^{1/2}$ relative to those from the stationary points.
The sum in the first line above is to be taken over all stationary points
$x_i$ which lie in the integration interval $a \leq x_i \leq b$. If either 
of the end points $a$ or $b$ is a stationary point, its contribution to 
the sum has to be divided by two and the corresponding term in the second 
line above has to be omitted. If there is no stationary point at all in 
the interval $[a,b]$, there is no contribution to the first line and the 
leading terms of the integral $I$ are of order $\hbar$ as given by the 
end-point contributions in the second line alone.

The stationarity condition for the phase in the integral \eq{dg2gs} leads to
the resonance condition for the rational tori
\be
k_v\,T_A(E_{k_u k_v}^*) = 2\pi k_u = k_u\, T_B\,.          \label{res}
\ee
Note that this condition is independent of the energy $E$. The stationary points 
$E_v=E_{k_u k_v}^*$ are the energies at which the $k_v$-th repetition of the A orbit 
bifurcates, cf. equations \eq{res1}, \eq{res2} in section 4.3. The condition \eq{res} 
can only be fulfilled if $k_u$ and $k_v$ have the same sign and if $|k_u|\geq |k_v|$. 
In most formulae below, we take $k_u$ and $k_v$ to be positive (or zero for one of 
them) and account for the two signs by an extra factor of two in the summations, 
taking real parts where necessary. For $k_u=k_v=k$, 
the stationary point is at $E_{kk}^*=0$, for all other tori the stationary
points are at finite energies. This gives, according to the first line in 
\eq{wongf}, the following asymptotic contribution to lowest order in $\hbar$:
\be
\delta g_{as}^{(T)}(E) = \sum_{k_v=1}^\infty \;\sum_{k_u=k_v}^\infty
                    {\cal A}_{T_{k_u k_v}}
                    \cos\left[\frac{1}{\hbar}\,S_{T_{k_u k_v}}(E) +
                    \frac{\pi}{4}\right]
                    \Theta(E-E_{k_u k_v}^*)\,,
\label{dgt}
\ee
which is exactly the Berry-Tabor trace formula \cite{beta,crli} with the
amplitudes
\be
{\cal A}_{T_{k_u k_v}} = f_{k_u k_v}
                         \frac{(-1)^{k_u+k_v}}{\hbar^{3/2}}\,\frac{k_u}{k_v}\,
                         \sqrt{\frac{2\pi}{k_vT_A'(E_{k_u k_v}^*)}}\,,\quad
           f_{k_u k_v} = \left\{ \begin{array}{c} 2 \\ 1 \end{array} \;
                         \hbox{for} \; \begin{array}{c} k_u\neq k_v \\k_u=k_v
                                       \end{array} \right\}.
\label{tamp} 
\ee
The actions of the tori are
\be
S_{T_{k_u k_v}}(E) = k_v\,S_A(E_{k_u k_v}^*)+2\pi k_u\,(E-E_{k_u k_v}^*)\,.
                     \qquad (E\geq E_{k_u k_v}^*)
\label{st}
\ee
The diagonal torus $T_{kk}$, which comes from the lower end point $E_v=0$
of the integral in \eq{dg2gs}, corresponds to the $k$-th repetition of the
classical B orbit. This is somewhat puzzling, since classically 
this orbit appears to be isolated along the $u$ axis, but semiclassically it 
contributes in the same way as the two-dimensional tori with $k_u\neq k_v$
with an amplitude proportional to $\hbar^{-3/2}$. The reason for this is
connected to the fact that the energy $E_{kk}^*=0$ at which it bifurcates is 
simultaneously the limit of the isotropic two-dimensional harmonic oscillator 
in which all orbits form a two-dimensionally degenerate family with SU(2) 
symmetry. The contributions from $k_v=0$ and $k_u\neq 0$ lead to a small 
correction
\be
\delta g_{as}^{(B0)}(E) =  \frac{1}{\pi\hbar}\sum_{k_u=1}^\infty 
                           \frac{(-1)^{k_u}}{k_u}
                           \sin\left(\frac{k_u}{\hbar}\,
                           2\pi E\right),                  \label{dgb0} 
\ee
which is of order $\sqrt{\hbar}$ with respect to \eq{tamp} and found to be
negligible in our numerical calculations.

The upper end point $E_v=E$ of the integral in \eq{dg2gs} corresponds to 
motion along the $v$ axis which classically gives the isolated A orbit. According 
to the second line in \eq{wongf}, this yields the asymptotic contribution
\be
\delta g^{(A)}_{as}(E) = \frac{T_A(E)}{\pi\hbar}\,
                         \sum_{k_v=1}^\infty \sum_{k_u=-\infty}^\infty
                         \frac{(-1)^{k_u+k_v}}{[k_vT_A(E)-2\pi k_u]}\,
                         e^{i[k_vS_A(E)/\hbar-\pi/2]}.            \label{dgak}
\ee
Using the identity \eq{polesum} we can do the summation over $k_u$ analytically
and find
\be
\delta g^{(A)}_{as}(E) = \frac{T_A(E)}{2\pi\hbar} \sum_{k_v=1}^\infty
                         \frac{(-1)^{k_v}}{\sin[k_vT_A(E)/2]\,}\,
                         \cos\left[\frac{k_v}{\hbar}\,S_A(E) -
                         \frac{\pi}{2}\right],                  \label{dga}
\ee
which is exactly the Gutzwiller trace formula for the isolated A orbit.
Of course, this expression cannot be used at the bifurcation energies of the
A orbit where, on one hand, \eq{polesum} is not valid and, on the other hand,
the upper end-point correction from the integral \eq{dg2gs} should be replaced
by one half of the corresponding torus contribution to \eq{dgt}. The 
contributions from $k_u=0$ and $k_v\neq 0$ in \eq{dgak} lead to a small 
correction
\be
\delta g_{as}^{(A0)}(E) = \frac{1}{\pi\hbar}\sum_{k_v=1}^\infty 
                          \frac{(-1)^{k_v}}{k_v}\,
                          \sin\left(\frac{k_v}{\hbar}\,
                          S_A(E)\right)                    \label{dga0}  
\ee
which is included in \eq{dgak}, \eq{dga} and will be referred to below.

We have thus established that the isolated A orbit emerges asymptotically, 
with its standard Gutzwiller amplitude \cite{rich}, from the upper end-point 
corrections of the EBK trace formula \eq{dg2gs}, whereas the tori with their 
standard Berry-Tabor amplitudes come from the stationary points of the phase 
in the integral \eq{dg2gs}. 

In order to obtain finite amplitudes at the bifurcations and the symmetry 
point $e=0$, we have to develop a uniform approximation. This can be done
quite easily by expanding the phase and the amplitude of the integrand in 
\eq{dg2gs} around the stationary points $E_{k_u k_v}^*$ up to first and 
second order in $E_v-E_{k_u k_v}^*$, respectively. Noting that the torus 
action $S_{T}(E)$ in \eq{st} represents the first two terms of the same 
expansion of $S_A(E)$ around $E-E_{k_u k_v}^*$, this leads to the approximate 
contributions
\be
\hspace{-1.2cm}  \Re e \left\{  e^{\frac{i}{\hbar}S_{T_{k_u k_v}}(E)
                     -i\pi(k_u+k_v)}\int_0^E
                     \left(b_{k_u k_v}+c_{k_u k_v}E_v\right)\,
                     e^{\frac{i}{2\hbar}\,a_{k_u k_v} (E_v-E_{k_u k_v}^*)^2}\,
                     \d E_v \right\}                            \label{gkuint}
\ee
which are exactly of the same type as those which we have derived from the
normal form theory in appendix B, and which can be integrated analytically
using the formulae \eq{j1} and \eq{j2undj3}. The parameters $a_{k_u k_v}$, 
$b_{k_u k_v}$ and $c_{k_u k_v}$ must be determined by the requirement that the 
asymptotic amplitudes and actions of both the A orbit and the T$_{k_u k_v}$ 
torus be recovered far away from the bifurcation energy $E_{k_u k_v}^*$. This 
procedure is completely analogous to what has been discussed in the earlier
appendices and need not be repeated here. The final uniform approximation which
we obtain after summing over all tori is
\bea 
\delta g_{uni}(E) \!\!& = & \!\!\!\sum_{k_v=1}^\infty\,\sum_{k_u=k_v}^\infty\!
(-1)^{k_u+k_v}\!\Biggl\{\!\Bigg({\cal A}_{A_{k_uk_v}}(E)-\frac{1}{2}\,\sigma_{k_uk_v}\,
 \sqrt{\frac{\hbar}{\pi\Delta S_{k_u k_v}}}\,{\cal A}_{T_{k_u k_v}}\!\Bigg)
\cos\left[\frac{k_v}{\hbar}\,S_A(E)-\frac{\pi}{2} \right] 
      \nonumber\\ 
&   & \hspace{1.6cm} \;+\;\; \frac{{\cal A}_{T_{k_u k_v}}}{2}\,\Re e \left[\left(
      e^{i\pi/4}[1-\delta_{k_uk_v}]
      +\sqrt{2}\,[C(\xi_{k_u k_v})+iS(\xi_{k_u k_v})]\right)
      e^{\frac{i}{\hbar}S_{T_{k_u k_v}}\!(E)}\right]\!\Bigg\}\nonumber\\
      \cr
& + & \!\!\delta g_{as}^{(A0)}(E) \;+\; \delta g_{as}^{(B0)}(E)\,, \label{dgugrap}
\eea
where we have included the small contributions \eq{dgb0} and \eq{dga0} to be 
consistent up to order $\hbar^{-1}$ in the amplitudes, although they are 
numerically insignificant. Here we have defined 
\be
{\cal A}_{A_{k_u k_v}}(E) = \frac{1}{\pi\hbar}\,
            \frac{2k_v\,[T_A(E)]^2}{[(k_vT_A(E))^2-(2\pi k_u)^2]}\,,\qquad
\sigma_ {k_uk_v}= {\rm Sign}(E-E_{k_u k_v}^*)\,,                  \label{aamp}
\ee
and 
\be
\xi_{k_u k_v} = \sigma_{k_uk_v}\,\sqrt{\frac{2\Delta S_{k_u k_v}}{\pi\hbar}}\,,
      \qquad \Delta S_{k_u k_v}(E) = k_v S_A(E)-S_{T_{k_u k_v}}(E)\geq 0\,,
\label{sigdel}
\ee
and the amplitudes and actions of the tori are given in \eq{tamp} and \eq{st},
respectively.

The uniform trace formula \eq{dgugrap} is discussed and tested versus
quantum-mechanical results in section 4.3. Here we compare its results 
with those of a numerical integration of the EBK trace formula \eq{dg2gs}.
We choose the value $\lambda=0.04$, where the saddle energy is $E=104.666$,
and a resolution of the energy spectrum limited by $|k_u|,|k_v|\leq 8$. In 
the figures \ref{dguni1} - \ref{dguni3}, covering different energy regions, 
the upper panels show the results of \eq{dg2gs} by the solid
lines and the sum of the asymptotic Berry-Tabor contributions of the tori
\eq{dgt} plus the Gutzwiller contribution \eq{dga} of the isolated A orbit
by the dashed lines. In the lower panel, the same results of \eq{dg2gs} 
(solid lines) are compared to those of the uniform approximation \eq{dgugrap}
(dashed lines). In all cases, the latter proves to be an excellent 
approximation to the exactly integrated EBK trace formula \eq{dg2gs}.

\Figurebb{dguni1}{65}{10}{950}{548}{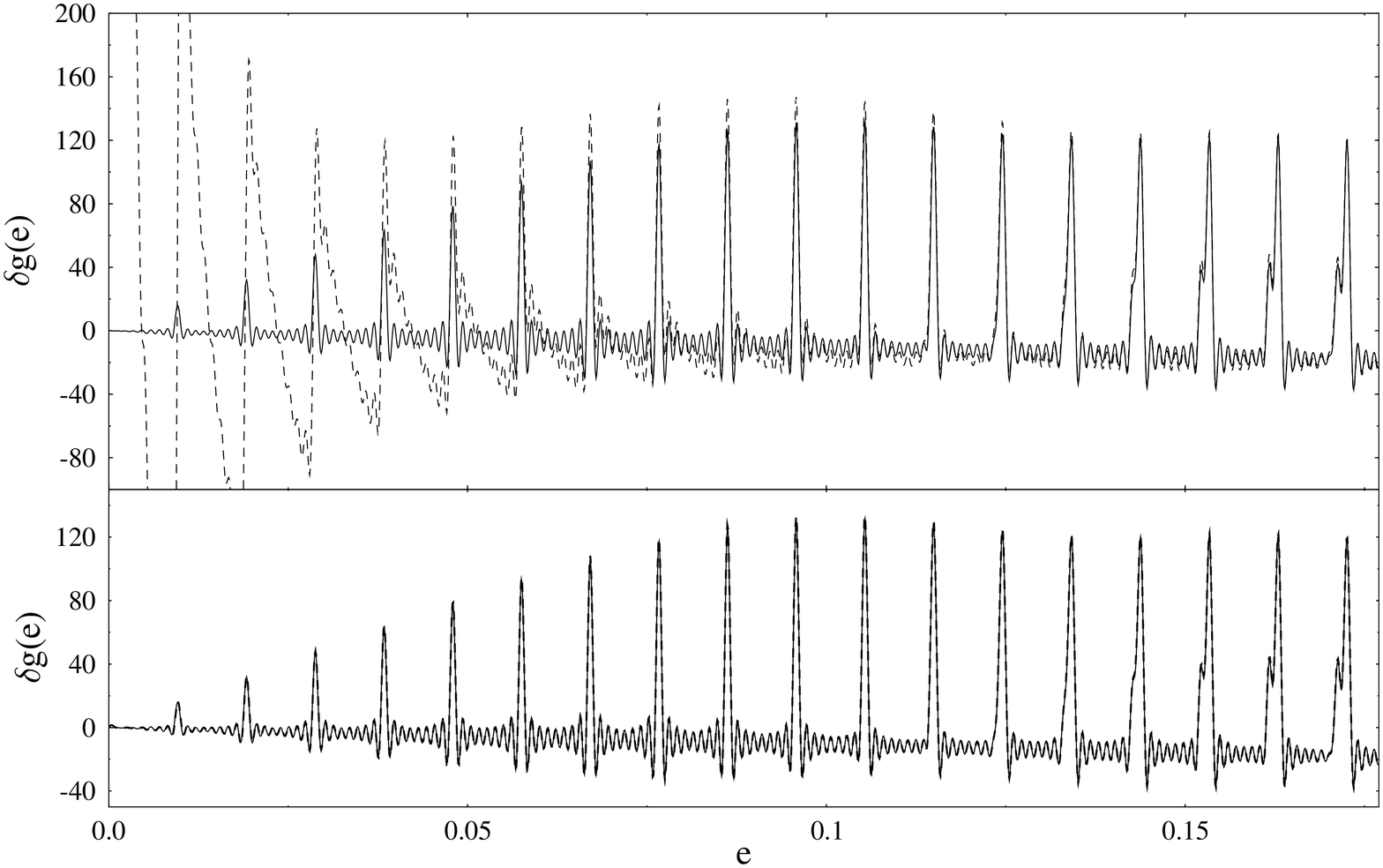}{10}{16.6}{
Shell structure in the level density of the integrable H\'enon-Heiles system.
{\it Solid lines:} exact EBK integral \eq{dg2gs}. {\it Dashed lines:} asymptotic 
approximations. {\it Upper panel:} Berry-Tabor contributions \eq{dgt} 
of the T tori plus Gutzwiller contribution \eq{dga} of the isolated A 
orbit. {\it Lower panel:} uniform approximation \eq{dgugrap}. In both cases,
repetition numbers $|k_u,k_v|\leq 8$ are included. Here the lowest energy
region is shown where the isolated A orbit contribution diverges in 
the limit $e \rightarrow 0$.
}

\vspace*{-0.5cm}
In \fig{dguni1} the lowest energy region is shown, where the Gutzwiller
contributions of the A orbit are seen to diverge in the limit $e\rightarrow 0$.
The divergences disappear in the uniform approximation. In \fig{dguni2}, an 
intermediate energy region is shown which includes the divergences
of the isolated A contributions at the bifurcations of the $k_u:k_v=9:8$ and
8:7 resonances. Figure \ref{dguni3} shows the top energy region containing
all resonances with $k_u:k_v\geq 3:2$.

These results demonstrate that the uniform approximation \eq{dgugrap}, which
expresses the level density in terms of the Berry-Tabor and Gutzwiller
amplitudes of the periodic orbits, reproduces the numerically integrated
EBK trace formula \eq{dg2gs} to a high degree of accuracy.

Finally we stress that our above derivation of the uniform approximation
\eq{dgugrap} is not limited to the separable H\'enon-Heiles system, but can
easily be modified to any separable potential by starting from the general
EBK trace formula \eq{gfold} rather than from \eq{dg2gs}.

\Figurebb{dguni2}{75}{15}{767}{535}{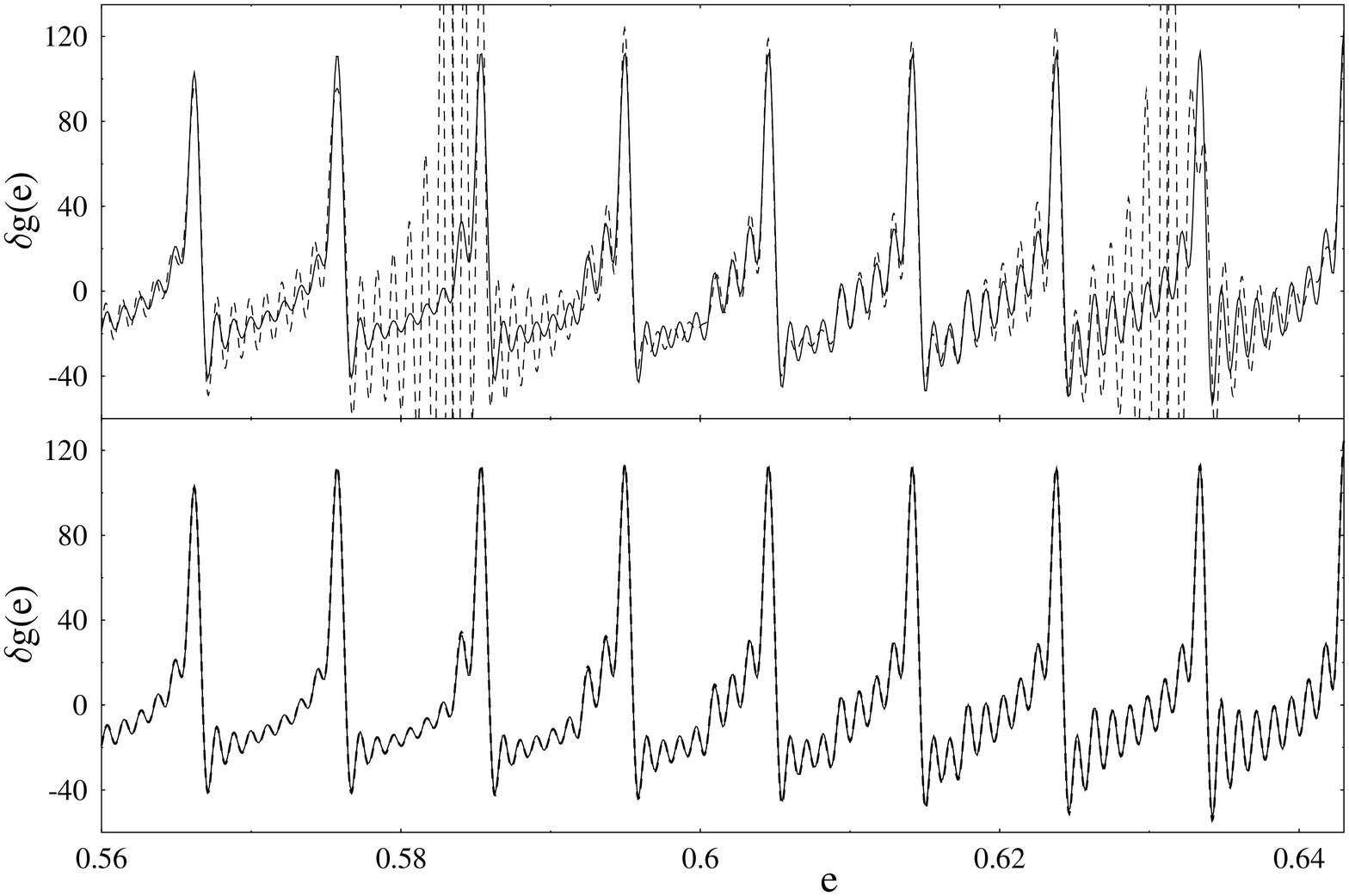}{10}{16.6}{
Same as \fig {dguni1} in an intermediate energy region. The top panel exhibits 
the divergence of the Gutzwiller contributions of the A orbit (dashed line) 
near the $k_u:k_v=9:8$ and 8:7 resonances. 
}
\vspace*{-1cm}
\Figurebb{dguni3}{75}{15}{767}{535}{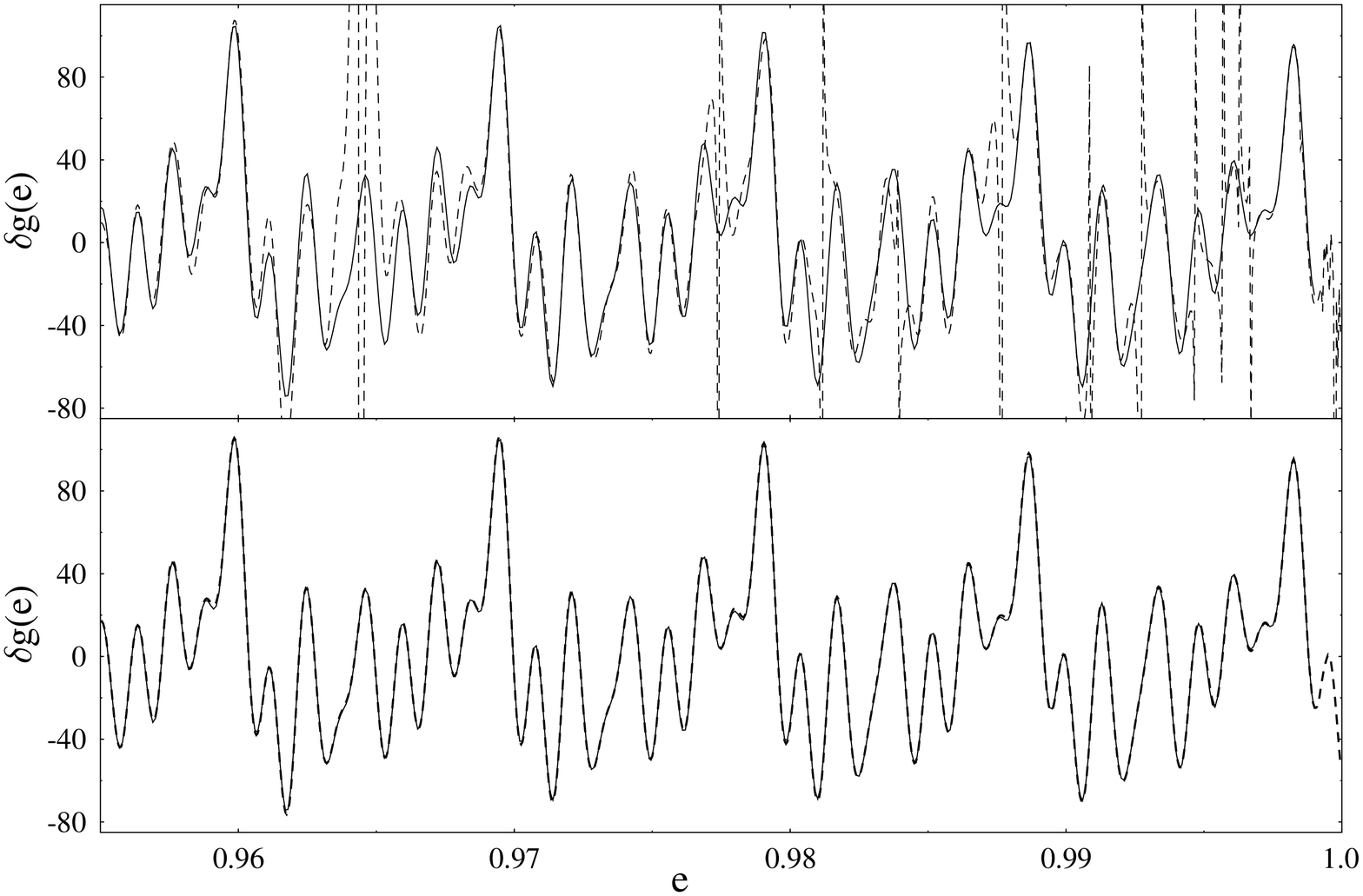}{10}{16.6}{
Same as \fig{dguni1} in the top energy range, covering the resonances with
$k_u:k_v\geq 3:2$.
}


\newpage

\begin{thebibliography}{10}

\bibitem{gutz71} M. C. Gutzwiller, J. Math.\ Phys.\ {\bf 12}, 343 (1971).

\bibitem{gutzbuch} M. C. Gutzwiller: {\it Chaos in Classical and 
                   Quantum Mechanics} (Springer, New York, 1990)

\bibitem{stma} V. M. Strutinsky and A. G. Magner,
               Sov.\ J. Part.\ Nucl.\ {\bf 7}, 138 (1976).

\bibitem{beta} M. V. Berry M V and M. Tabor, Proc.\ R. Soc.\ A {\bf 
               349}, 101 (1976);\\
               M. V. Berry and M. Tabor, J. Phys.\ A {\bf 10}, 371 (1977).

\bibitem{bablo}  R. Balian and C. Bloch, Ann.\ Phys.\ (N. Y.) {\bf 69},
                 76 (1972).

\bibitem{sm77} V. M. Strutinsky, A. G. Magner, S. R. Ofengenden, and T. 
               D{\o}ssing, Z. Phys.\ A {\bf 283}, 269 (1977).

\bibitem{crli} S. C. Creagh and R. G. Littlejohn, Phys.\ Rev.\ 
               A {\bf 44}, 836 (1991);\\
               S. C. Creagh and R. G. Littlejohn, J. Phys.\ A 
               {\bf 25}, 1643 (1992).

\bibitem{ozoha} A. M. Ozorio de Almeida and J. H. Hannay, J. Phys.\ A 
                {\bf 20} 5873 (1987).

\bibitem{ozobu} A. M. Ozorio de Almeida: {\it Hamiltonian Systems: Chaos 
                and Quantization} (Cambridge University Press, 1988).

\bibitem{crper} S. C. Creagh, Ann.\ Phys.\ (N.Y.) {\bf 248}, 60 (1996).

\bibitem{toms} S. Tomsovic, M. Grinberg and D. Ullmo, Phys.\ Rev.\ 
               Lett.\ {\bf 75}, 4346 (1995);\\
               D. Ullmo, M. Grinberg and S. Tomsovic, Phys.\ Rev.\ E 
               {\bf 54}, 135 (1996).

\bibitem{hhun} M. Brack, P. Meier and K. Tanaka, J. Phys.\ A
               {\bf 32}, 331 (1999).

\bibitem{si96} M. Sieber, J. Phys.\ A {\bf 29}, 4715 (1996).

\bibitem{ss97} H. Schomerus and M. Sieber, J. Phys.\ A 
               {\bf 30} 4537 (1997).

\bibitem{ss98} M. Sieber and H. Schomerus, J. Phys.\ A 
               {\bf 31}, 165 (1998).

\bibitem{maodel} J.-M. Mao and J. B. Delos, Phys.\ Rev.\ A {\bf 45}, 
                 1746 (1992).

\bibitem{main} J. Main, G. Wiebusch, A. Holle and K. H. Welge, 
               Phys.\ Rev.\ Lett.\ {\bf 57} 2789 (1996).

\bibitem{mbgu} M. Brack in: {\it Festschrift in honor of the 75th birthday 
               of Martin Gutzwiller}, Eds. A. Inomata \etal; 
               Foundations of Physics {\bf 31}, 209 (2001). 
               [LANL preprint nlin.CD/0006034]

\bibitem{lamp} M. Brack, M. Mehta and K. Tanaka, J. Phys.\ A 
               {\bf 34}, 8199 (2001).  

\bibitem{meyer} K. R. Meyer, Trans.\ Am.\ Math.\ Soc.\ {\bf 149}, 95 (1970).

\bibitem{bruno} A. D. Bruno, Math.\ USSR Sbornik {\bf 12}, 271 (1970);\\
                A. D. Bruno, Inst.\ Prikl.\ Mat.\ Akad.\ Nauk SSSR 
                Preprint No.\ 18 (1972) (Moscow, in Russian)

\bibitem{then} H. Then, Diploma thesis (University of Ulm, 1999);\\
               H. Then and M. Sieber, to be published.

\bibitem{sc97} H. Schomerus, Europhys.\ Lett.\ {\bf 38}, 423 (1997);\\
               H. Schomerus and F. Haake, Phys.\ Rev.\ Lett.\
               {\bf 79}, 1022 (1997).

\bibitem{sc98} H. Schomerus, J. Phys.\ A {\bf 31}, 4167 (1998).

\bibitem{sadov} D. A. Sadovskii, J. A. Shaw and J. B. Delos,
                Phys.\ Rev.\ Lett.\ {\bf 75}, 2120 (1995);\\
                D. A. Sadovskii and J. B. Delos, Phys.\ Rev.\ E 
                {\bf 54}, 2033 (1996).

\bibitem{feig} M. J. Feigenbaum, J. Stat.\ Phys.\ {\bf 19}, 25 (1978);\\
               see also M. J. Feigenbaum, Physica {\bf 7 D}, 16 (1983).

\bibitem{hhpaper} M. H\'enon and C. Heiles, Astr.\ J. {\bf 69}, 73 (1964).

\bibitem{bermoun} M. V. Berry and K. E. Mount, Rep.\ Prog.\ Phys.\ {\bf 35}, 
                  315 (1972).

\bibitem{book} M. Brack and R. K. Bhaduri: {\it Semiclassical Physics}, 
               Frontiers in Physics Vol.\ 96 (Westview Press, Bolder, 2003).

\bibitem{strut} V. M. Strutinsky, Nucl.\ Phys.\ A {\bf 122}, 1 (1968). 

\bibitem{abro} M. Abramowitz and I. A. Stegun: {\it Handbook 
               of Mathematical Functions}, 9th printing (Dover, New York, 1970)

\bibitem{richens} P. J. Richens J. Phys.\ A {\bf 15}, 2110 (1982).

\bibitem{chur} R. C. Churchill, G. Pecelli and D. L. Rod in: 
               {\it Stochastic Behavior in Classical and Quantum 
               Hamiltonian Systems}, Eds. G. Casati and J. Ford  
               (Springer-Verlag, New York, 1979) p.\ 76.

\bibitem{davies} K. T. R. Davies, T. E. Huston and M. Baranger, Chaos
                 {\bf 2}, 215 (1992).

\bibitem{hhprl} M. Brack, R. K. Bhaduri, J. Law and M. V. N. Murthy, Phys.\ 
                Rev.\ Lett.\ {\bf 70}, 568 (1993);\\
                M. Brack, R. K. Bhaduri, J. Law, M. V. N. Murthy and Ch. Maier,
                Chaos {\bf 5}, 317 and 707(E) (1995).
 
\bibitem{jkpw} J. Kaidel and P. Winkler, to be published.

\bibitem{erde} A. Erd\'elyi: {\it Asymptotic expansions} (Dover, New York, 1956).

\bibitem{wong} R. Wong: {\it Asymptotic Approximation of Integrals}
               (Academic Press, San Diego, 1989).

\bibitem{brja} M. Brack and S. R. Jain, Phys.\ Rev.\ A {\bf 51}, 3462 (1995).

\bibitem{rich} The same result was obtained by P. J. Richens, J. Phys.\ A 
               {\bf 15}, 2101 (1982), starting from the action-angle variables
               for a general two-dimensional integrable system containing an 
               isolated orbit.

\end{thebibliography}
\end{document}